\documentclass[12pt,preprint]{aastex}

\def\cgs{erg cm$^{-2}$ s$^{-1}$} 
\def\gsimeq{\hbox{\raise0.5ex\hbox{$>\lower1.06ex\hbox{$\kern-1.07em{\sim}$}$}}} 
\def\lsimeq{\hbox{\raise0.5ex\hbox{$<\lower1.06ex\hbox{$\kern-1.07em{\sim}$}$}}} 

\shorttitle{$z>3$ QSOs in XMM-COSMOS}
\shortauthors{Brusa et al. }

\begin{document}


\title{High redshift quasars in the COSMOS survey: the space density of $z>$ 3 
X-ray selected QSOs}


\author{M. Brusa\altaffilmark{1}, A. Comastri\altaffilmark{2}, R. Gilli\altaffilmark{2}, G. Hasinger\altaffilmark{1},
  K. Iwasawa\altaffilmark{2,1}, V. Mainieri\altaffilmark{3}, M. Mignoli\altaffilmark{2}, M. Salvato\altaffilmark{4},
  G. Zamorani\altaffilmark{2}, A. Bongiorno\altaffilmark{1},  N. Cappelluti\altaffilmark{1}, F. Civano\altaffilmark{5},
  F. Fiore\altaffilmark{6}, A. Merloni\altaffilmark{1,7}, J. Silverman\altaffilmark{8}, J. Trump\altaffilmark{9},
  C. Vignali\altaffilmark{10}, P. Capak\altaffilmark{4}, M. Elvis\altaffilmark{5}, O. Ilbert\altaffilmark{11}, 
  C. Impey\altaffilmark{9}, S. Lilly\altaffilmark{8}}

\altaffiltext{1}{Max Planck Institut f\"ur Extraterrestrische Physik,
  Giessenbachstrasse, 1, D-85748, Garching bei Muenchen, Germany}
\altaffiltext{2}{INAF --  Osservatorio Astronomico di Bologna, via Ranzani 1, 
I--40127 Bologna, Italy}
\altaffiltext{3}{European Southern Observatory, Karl-Schwarzschild-str. 2,
  85748 Garching bei M\"unchen, Germany} 
\altaffiltext{4}{California Institute of Technology, MC 105-24, 1200 East
California Boulevard, Pasadena, CA 91125} 
\altaffiltext{5}{Harvard-Smithsonian Center for Astrophysics, 60 Garden
Street, Cambridge, MA 02138 }
\altaffiltext{6}{INAF --  Osservatorio Astronomico di Roma, via Frascati 33,
Monteporzio-Catone (Roma), I-00040, Italy}
\altaffiltext{7}{Excellence Cluster Universe, 
  Boltzmannstrasse 2, D-85748, Garching bei Muenchen, Germany}
\altaffiltext{8}{Institute of Astronomy, Swiss Federal Institute of  
Technology (
ETH H\"onggerberg), CH-8093, Z\"urich, Switzerland.}
\altaffiltext{9}{Steward Observatory, University of Arizona, 933 North
Cherry Avenue, Tucson, AZ 85 721}
\altaffiltext{10}{Dipartimento di Astronomia Universit\'a di Bologna, via Ranzani 1, 
I--40127 Bologna, Italy}
\altaffiltext{11}{Univ. Hawaii, 2680 Woodlawn Dr., Honolulu, HI, 96822}


\begin{abstract}
We present a new measurement of the space density of high redshift 
($z \simeq $ 3.0-4.5), X--ray selected QSOs obtained by exploiting the 
deep and uniform multiwavelength coverage of the COSMOS survey. 
We have assembled a statistically large (40 objects), X--ray selected 
($F_{0.5-2 keV} >$  10$^{-15}$ \cgs), homogeneous sample of $z >$ 3  QSOs
for which spectroscopic (22) or photometric (18) redshifts are available. 
We present the optical (color-color diagrams) and 
X--ray properties, the number counts and space densities 
of the $z >$ 3 X--ray selected quasars population 
and compare our findings with previous works and model predictions.  
We find that the optical properties of X--ray selected quasars are not
significantly different from those of optically selected samples.
There is evidence for substantial X--ray absorption (log$N_H >$ 23 cm$^{-2}$) 
in about 20\% of the sources in the sample. 
The comoving space density of luminous ($L_X \gsimeq 
10^{44}$ erg s$^{-1}$) QSOs declines exponentially (by an e--folding  
per unit redshift) in the $z \sim$ 3.0--4.5 range, with a behavior similar
to that observed for optically bright unobscured QSOs selected in large area
optical surveys. 
Prospects for future, large and deep X--ray surveys are also discussed.

\end{abstract}



\section{Introduction}

The well known correlations between Super Massive Black Holes (SMBH) and galaxy properties 
such as bulge luminosity and stellar velocity dispersion 
(Gebhardt et al. 2000; Ferrarese \& Merrit 2000), which are by now 
firmly established at least in the local Universe, point 
towards a tight link between the assembly of the bulge mass and SMBH. 
The high redshift AGN luminosity function (LF) and the source counts represent
key observational constraints for theoretical models of galaxy and SMBH 
formation and evolution. While most models are reasonably successful 
in reproducing several observables in the high-luminosity regime 
(Di Matteo et al. 2005; Volonteri \& Rees 2006; 
Hopkins et al. 2005; Lapi et al. 2006; Menci et al. 2008), 
they have to face the paucity of data, 
especially at high redshifts and low luminosities.
Given that theoretical predictions are used to determine 
key physical parameters, such as the QSOs duty cycle, the BH seed mass function, 
and the accretion rates,  a reliable observational estimate of the 
QSOs LF and evolution at high redshift is extremely important. 

Large optical surveys, most notably the 2dF Quasar Redshift Survey 
(2QZ; Croom et al. 2004) and the Sloan Digital Sky Survey (SDSS; i.e. Richards et al. 2006)
were able to accurately measure the shape of the LF up to 
$z \sim 2.0-2.5$ and to constrain the 
bright (M$_{\rm  B}<$ --25.5 or M$_{\rm  I}<$ --27.6, corresponding to
bolometric luminosities of log$L_{bol}\gsimeq 46$ erg s$^{-1}$) QSOs LF up to z $\sim$
6.5:  
the space density of optically selected QSOs peaks at 
$z \sim 2-3$, then decreases approximately 
by a factor 3, per unit redshift, in the range $z\simeq 3-6$ 
(Schmidt, Schneider \& Gunn 1995, Fan et al. 2001, 2004; Richards et
al. 2006). 
The high--redshift quasar surveys mentioned above are relatively shallow 
and probe only the bright end of the LF. Recently, deep optical surveys 
were able to probe the LF at significantly fainter magnitudes 
using different selection criteria (e.g. Wolf et al. 2003, Hunt et
al. 2004, Fontanot et al. 2007, Bongiorno et al. 2007,
Siana et al. 2008). However, the statistics on the faint QSOs population 
at $z>3$ 
is still limited  given the typically small area surveyed. 

Moreover, optically selected quasars are known to make up only a 
fraction of the entire population of accreting SMBH: in fact, most of the 
accretion power in the Universe is obscured by large amounts of dust and gas
(see, e.g., Fabian \& Iwasawa 1999); moreover, AGN and QSOs span a broad range
in accretion rate (see, e.g., Merloni \& Heinz 2008), and, therefore, 
of intrinsic luminosities.  
Deep X-ray surveys with {\it Chandra} and XMM--{\it Newton} have proven to be 
very efficient in revealing obscured accretion (except for Compton Thick AGN) 
down to relatively low luminosities (see Brandt \& Hasinger 2005 for a
review).  
According to the most recent models for the synthesis of the X--ray Background 
(e.g. Gilli,Comastri \& Hasinger 2007), the obscured AGN  
population (including heavily obscured Compton Thick AGN) outnumbers the unobscured ones
by a luminosity dependent factor ranging from $\sim2$ at L$_{\rm X} \simeq 10^{45}$
erg s$^{-1}$ to $\sim8$ at L$_{\rm X} \simeq 10^{42}$ erg s$^{-1}$.
The evidence for a decreasing obscured AGN fraction towards high luminosities
is supported by many investigations
(e.g. Ueda et al. 2003, Treister \& Urry 2005). 
Moreover, there is also increasing evidence that the fraction of obscured AGN 
grows towards high redshifts (La Franca et al. 2005; Treister \& Urry
2006;  Hasinger 2008). 

By combining deep and shallower X--ray surveys (Ueda et al. 2003; La
Franca et  al. 2005; Hasinger et al. 2005; Silverman et al. 2008)  
it has been possible to address the issue of the evolution of the X-ray
LF at high redshifts. Previous works presented evidence for a
decline of the comoving space density of X--ray selected AGN at $z >$ 3 in
both  the soft (0.5--2 keV; Hasinger et al. 2005; Silverman et al. 2005) and hard (2--10 keV; Silverman 
et al. 2008) bands with a rate similar to that observed for optically selected quasars. 
At face value, these findings would imply that the space density of obscured AGN, 
which are likely to be missed by optical surveys, declines toward high redshifts ($z>$ 3) 
with a behavior similar to that of the unobscured AGN,
at least at the relatively high luminosities probed by present X--ray surveys
(log$L_{X}\gsimeq 44$).
However, the results available so far are based on rather heterogeneous 
and relatively small samples, also affected by a significant
level of spectroscopic incompleteness due to the faint magnitudes of the 
optical counterparts.
To cope with this limitation, several photometric selection techniques have
been proposed  
in the recent years. Though differing in the details, the search for high 
redshift, obscured AGN is mainly based on a suitable combination of X--ray, 
optical and infrared flux ratios (see e.g. Fontanot et al. 2007).
More specifically, hard X--ray emission associated with extremely faint or 
even undetected optical counterparts and extremely red colors (from optical 
wavelengths up to mid--infrared) is considered a reliable proxy 
for a high redshift obscured AGN (see, e.g., Koekemoer et al. 2004; Brusa et al. 2005; 
Fiore et al. 2008).  
As an example, at the X-ray fluxes probed by the COSMOS survey, an X-ray
  to optical 
flux ratio f$_X$/f$_{opt}>10$ efficiently selects high z, Compton thin obscured AGN (see Fiore et al. 2003).
Most of them have red or extremely red (R-K$>5$) optical to near infrared colors 
and indeed f$_X$/f$_{opt}$ and R-K are tightly correlated 
among X-ray selected sources (Brusa et al. 2005).
Unfortunately, due to the widely different selection criteria, it is difficult
to combine the above mentioned approaches to obtain a reliable estimate of their space density.
Given the importance of the AGN LF at high redshift 
for our understanding of SMBH growth and evolution, the search for and the census of
high redshift ($z >$ 3) AGN warrant further efforts.
The excellent multiwavelength coverage of the COSMOS field (Scoville et
al. 2007) offers the unique opportunity to assemble the first statistically
large, homogeneous, and  well defined sample of X--ray selected, high
redshift AGN, which is almost completely unbiased against obscuration up to
column densities of the order of $N_H \simeq 10^{23}$ cm$^{-2}$. 

The paper is structured as follows. In Section 2 we describe the sample
selection  which is mainly based on the optical and multiwavelength
identification  of the XMM--COSMOS survey (Hasinger et al. 2007, Cappelluti et
al. 2007, 2008, Brusa et al. 2007; 2008), complemented  by {\it Chandra}
positions when available (Elvis et al. 2008, Puccetti et al. 2008, Civano et
al. 2008) to secure the correct  optical counterpart, and, most 
importantly, by spectroscopic (Trump et al. 2008, Lilly et al. 2007) and
high-quality photometric  redshifts (Salvato et al. 2008). The 
optical and 
X--ray properties are presented in Section 3 and 4, respectively. In Section 5
the number counts and space density of the sample are discussed and compared
with model predictions. The results are summarized in Section 6, where the
perspectives for future observations are also briefly outlined. A $H_0 = 70$
km s$^{-1}$ Mpc$^{-1}$, $\Omega_M$ = 0.3 and  $\Omega_{\Lambda}$ = 0.7
cosmology is adopted. Optical magnitudes are in the AB system.

\section{The sample} 

\subsection{Parent sample} 

The Cosmic Evolution Survey (COSMOS) field (Scoville et al. 2007) is a so far
unique area for deep and wide comprehensive multiwavelength coverage, from the
optical band with {\it Hubble}, {\it Subaru} and other ground based telescopes, and infrared 
with {\it Spitzer}, and X-rays with XMM--{\it Newton} and {\it Chandra}, to the
radio with the VLA. 
The spectroscopic coverage with VIMOS/VLT and IMACS/Magellan, coupled with the
reliable photometric redshifts derived from multiband fitting,  
is designed to be directly comparable, at $0.5<z<1.0$, to the 2dFGRS (Colless et
al. 2001) at $z\sim0.1$, and to probe the correlated evolution of galaxies,
star formation, AGN, and dark matter with large-scale structure in the
redshift range $z \sim$ 0.5--4.    

The COSMOS field has been observed with XMM-{\it Newton} 
for a total of $\sim 1.5$ Ms  at a rather  homogeneous depth  
of $\sim 50$ ks (Hasinger et al. 2007, Cappelluti et al. 2007, 2008).
The catalogue used in this work includes 1848 point--like sources\footnote{In
  the present analysis we used 53 out of the 55 XMM-fields available; for this
  reason the number of sources is slighlty lower than that discussed in
  Cappelluti et al. (2008)} 
above a given threshold with a maximum likelihood detection algorithm in at
least one of the  soft (0.5--2 keV), hard (2--10 keV) or ultra-hard (5--10
keV) bands down to limiting fluxes of $\sim$ 5$\times 10^{-16}$, $\sim$ 3$\times 10^{-15}$  
and $\sim$ 5$\times 10^{-15}$ erg cm$^{-2}$ s$^{-1}$, respectively
(see Cappelluti et al. 2007, 2008 for more details). 
The adopted likelihood threshold corresponds to a probability $\sim
4.5\times10^{-5}$ that a catalog source is a spurious background fluctuation.
Following Brusa et al. (2008), we further excluded from this catalog 24
sources which turned out to be  a blend of two {\it Chandra} sources and
additional 26 faint XMM sources coincident with diffuse emission
(Finoguenov et al. 2008). 
To maximize the completeness over a well defined large area and, at the same
time, keep selection effects under control, we considered only sources detected 
above the limiting flux corresponding to a coverage of more than 1 deg$^{2}$
in at least one  of the three X--ray energy ranges considered, namely:
1$\times 10^{-15}$ \cgs, 6$\times 10^{-15}$
\cgs, and 1$\times 10^{-14}$ \cgs, in the 0.5--2 keV, 2--10 keV or 5--10 keV
bands, respectively (Cappelluti et al. 2008). 
The final sample includes therefore 1651 X--ray sources. 
The combination of area and depth is similar to the one of the sample 
 studied by Silverman et al. (2008,
  see their Fig. 8). The main difference is given by the considerably higher redshift
  completeness of the parent sample obtained thanks to the much deeper
  coverage in the optical and near IR bands and the systematic use of
  photometric redshifts (see below) to avoid the need for substantial
  corrections 
  for completeness (see discussion in Silverman et al. 2008). 

A detailed X--ray to optical association has been
performed (Brusa et al. 2008), applying the likelihood ratio technique
on the optical, near-infrared (K--band) and mid--infrared 
(IRAC) catalogs available. In addition, for 
the subsample of  the XMM-COSMOS sample covered by {\it Chandra}
observations ($\sim$ 50\% of the total sample), the {\tt ACIS} images
have been visually inspected to further exclude possible 
misidentifications problems.  
Of the 1651 sources in the XMM--COSMOS catalog described above, 
1465 sources have a unique/secure optical counterpart
from the multiwavelength analysis, i.e. among them we expect 
only $\sim$ 1\% of possible misidentifications (see discussion in Brusa et
al. 2008). 
For additional 175 sources, the proposed optical counterpart 
has about 50\% probability to be the correct one. 
Given that the alternative counterparts 
of these 175 sources show optical to IR properties and
redshift distribution comparable to the primary ones,
the proposed ones can be considered statistically representative of
the true counterparts of the X-ray sources. Therefore we consider also
those in our analysis. 
Finally, 11 sources remain unidentified, because 
it was not possible to assign them to any optical and/or infrared counterpart.

Spectroscopic redshifts for the proposed counterparts are available from the
Magellan/IMACS and MMT observation campaigns ($\sim 590$ objects, \citealt{trump},
Trump et al. 2008), from the zCOSMOS project ($\sim 350$ objects, \citealt{lilly}),
 or were already present either in the SDSS survey catalog 
($\sim 100$ objects, \citealt{sdss}, Kauffman et al. 2003\footnote{These 
  sources have been retrieved from the NED, NASA Extragalactic Database and
  from the SDSS archive}), 
or in the literature ($\sim 95$ objects, Prescott et al. 2006).  
In summary a total of 683 independent, good quality spectroscopic redshifts are
available corresponding to a substantial fraction ($\sim 40$\%) of 
the entire sample.

Photometric redshifts for all the XMM--COSMOS sources have been obtained 
exploiting the COSMOS multiwavelength database and are presented in
Salvato et al. (2008). 
Since the large majority of the XMM--COSMOS sources are AGN, in addition to the standard photometric
redshift treatments for normal galaxies, a new set
of SED templates has been adopted, together with a correction for long--term
variability and  luminosity priors for point-like sources (see Salvato et
al. 2008 for further details). 
Moreover, the availability of the intermediate band {\it Subaru} filters
(Taniguchi et al. 2008) is crucial in picking up emission lines (see also Wolf et al. 2004).
This led, for the first time for an AGN sample, to a photometric
redshift  accuracy comparable to that achieved for inactive
galaxies ($\sigma_{\Delta z/(1+z)} \sim 0.015$ and $\sim 5$\% outliers)
at i$\lsimeq$22.5. At fainter magnitudes (22.5 $<I<24.5$) the dispersions
increases to $\sigma_{\Delta z/(1+z)} \simeq 0.035$ with $\sim$ 15\% 
outliers, still remarkably good for an AGN sample. 

A photometric redshift is available for all but 36 objects out of 
1651 in the flux limited sample. 
Fourteen of them do not have multiband photometry, being detected only
in the IRAC and K bands. The remaining 22 objects are affected by severe
blending problems making the photo-z estimate not reliable. 

\begin{figure}
\includegraphics[width=17cm]{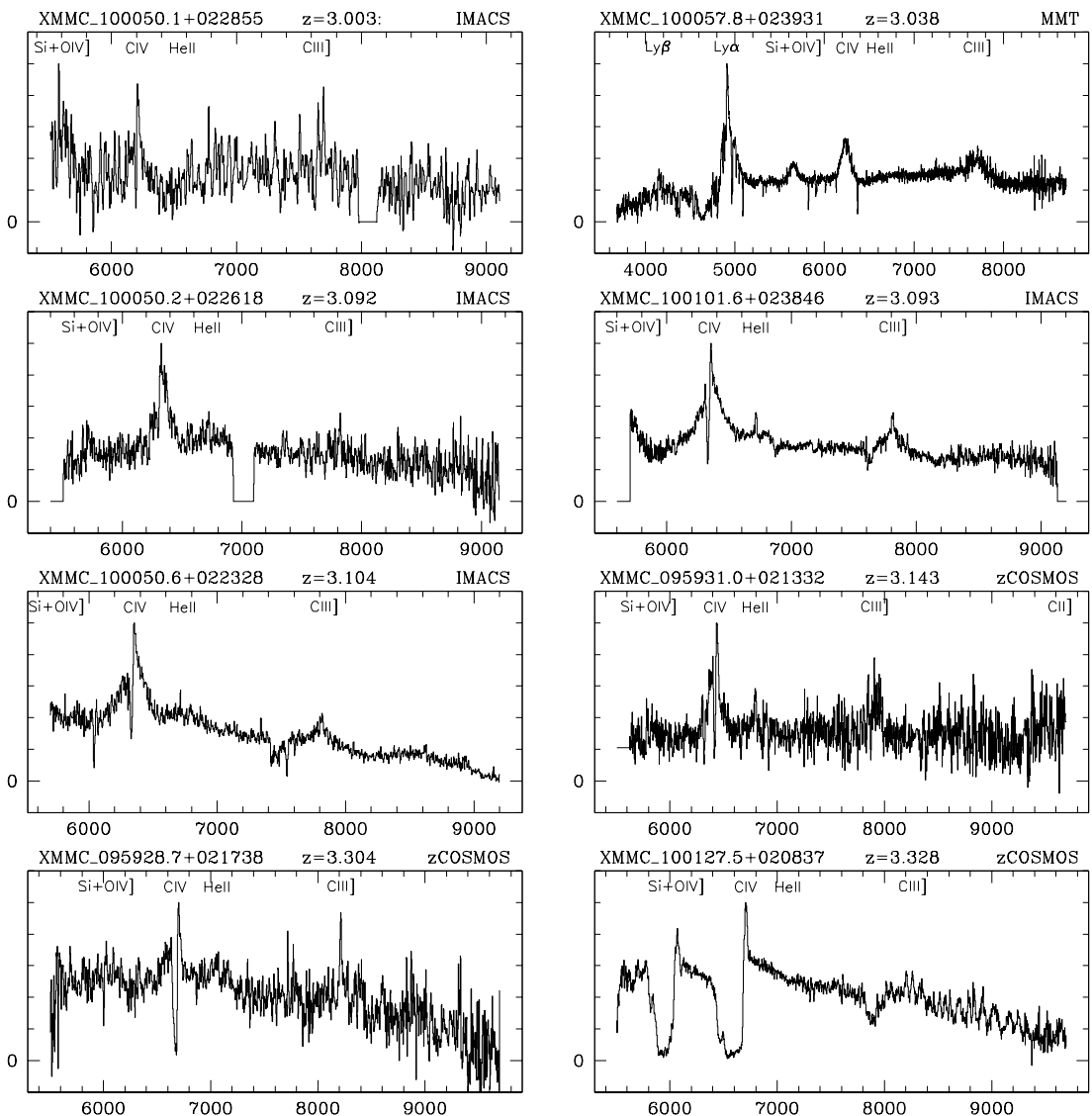}
\caption{
Spectra of the 22 spectroscopically confirmed z$>3$ quasars, in order of increasing 
redshift. The spectra are shown in the observed frame and the main emission
and absorption features are labelled. The four objects with single line
spectra are marked with a ``:'' right to the redshift. The source of the
spectra (IMACS, zCOSMOS or MMT is reported in the top right of each panel).}
\end{figure}
\begin{center}
\includegraphics[width=17cm]{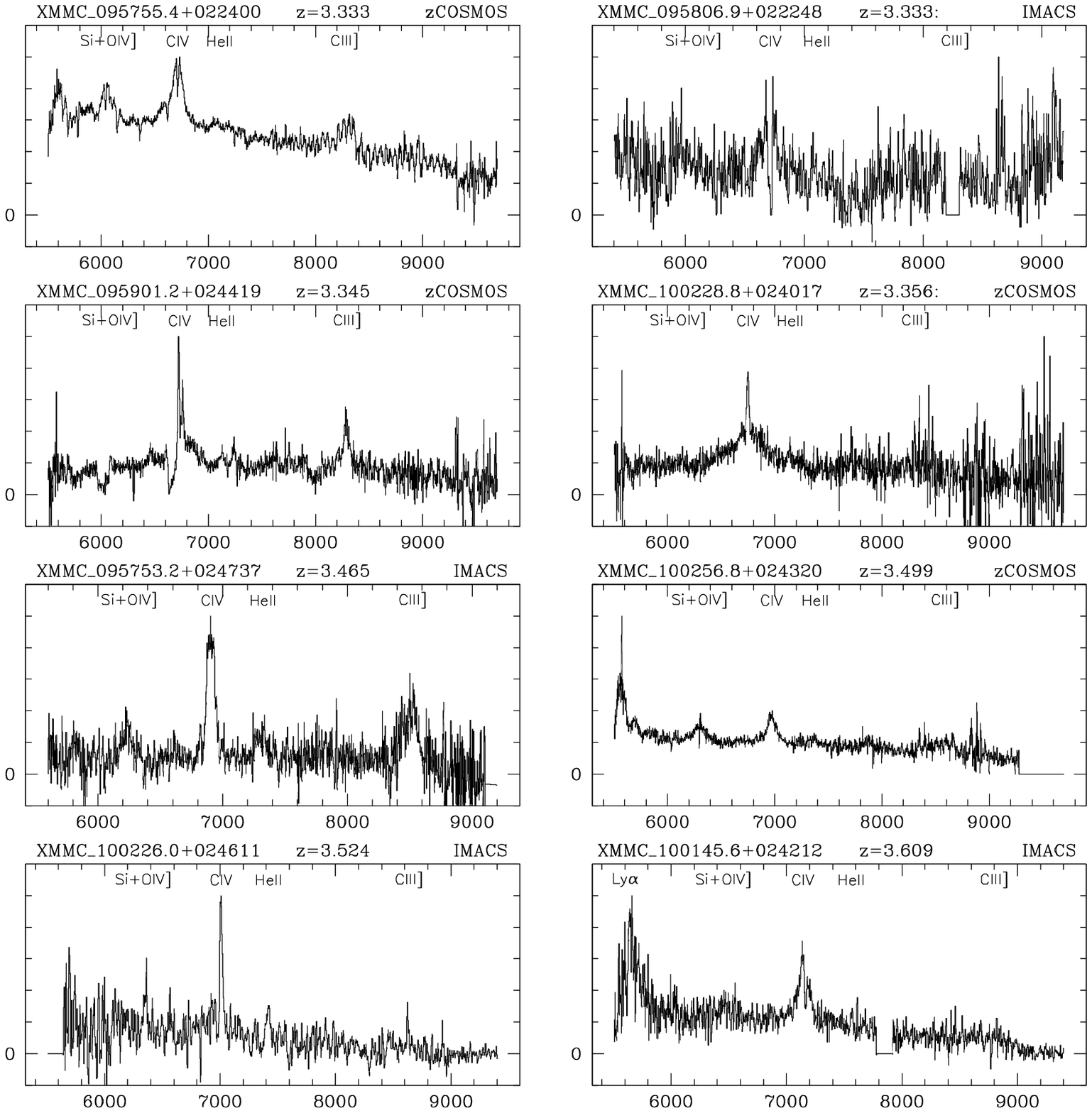}\\
{Fig. 1. --- Continued.}
\includegraphics[width=17cm]{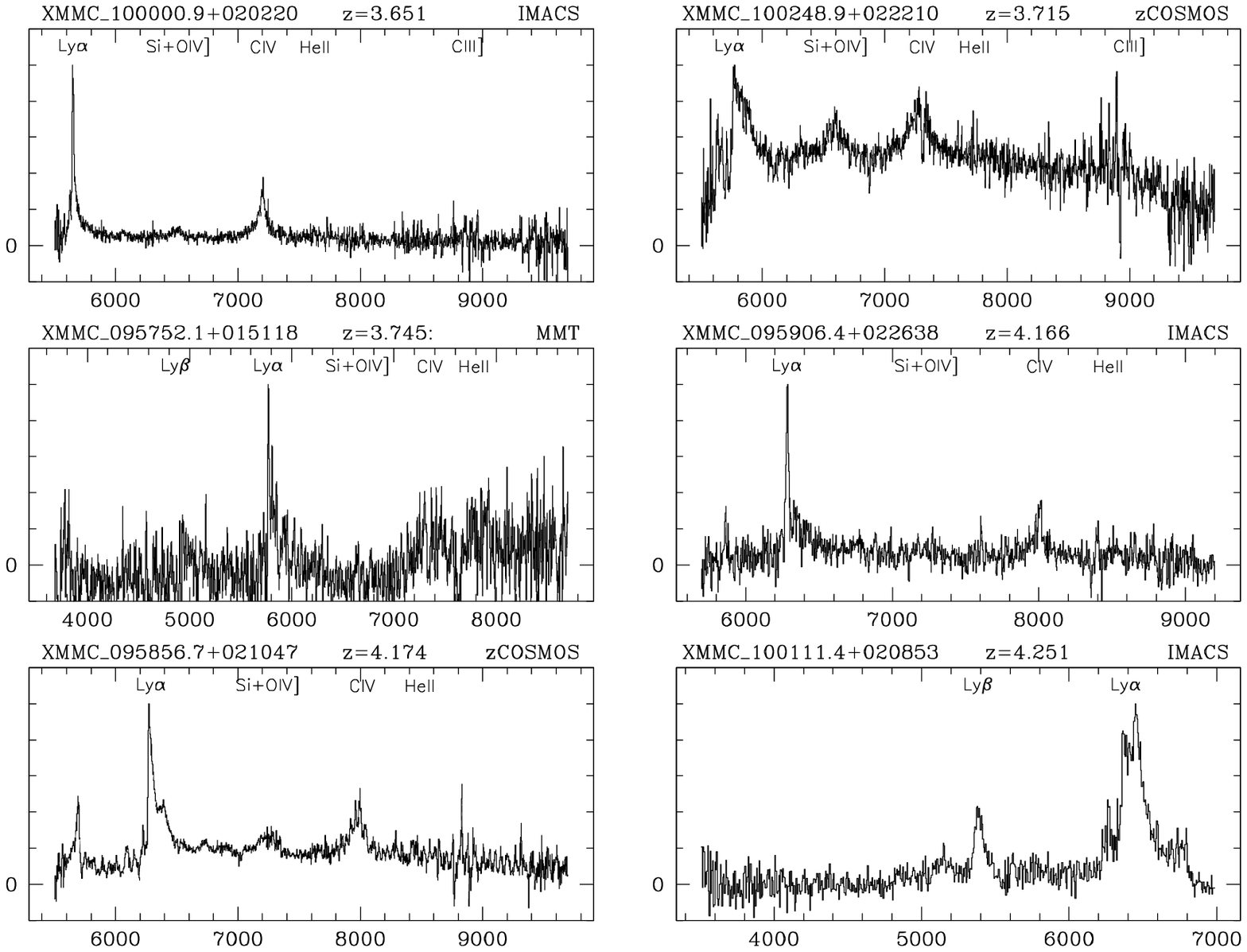}\\
{Fig. 1. --- Continued.}
\end{center}

\begin{figure}[!t]
\includegraphics[width=8cm]{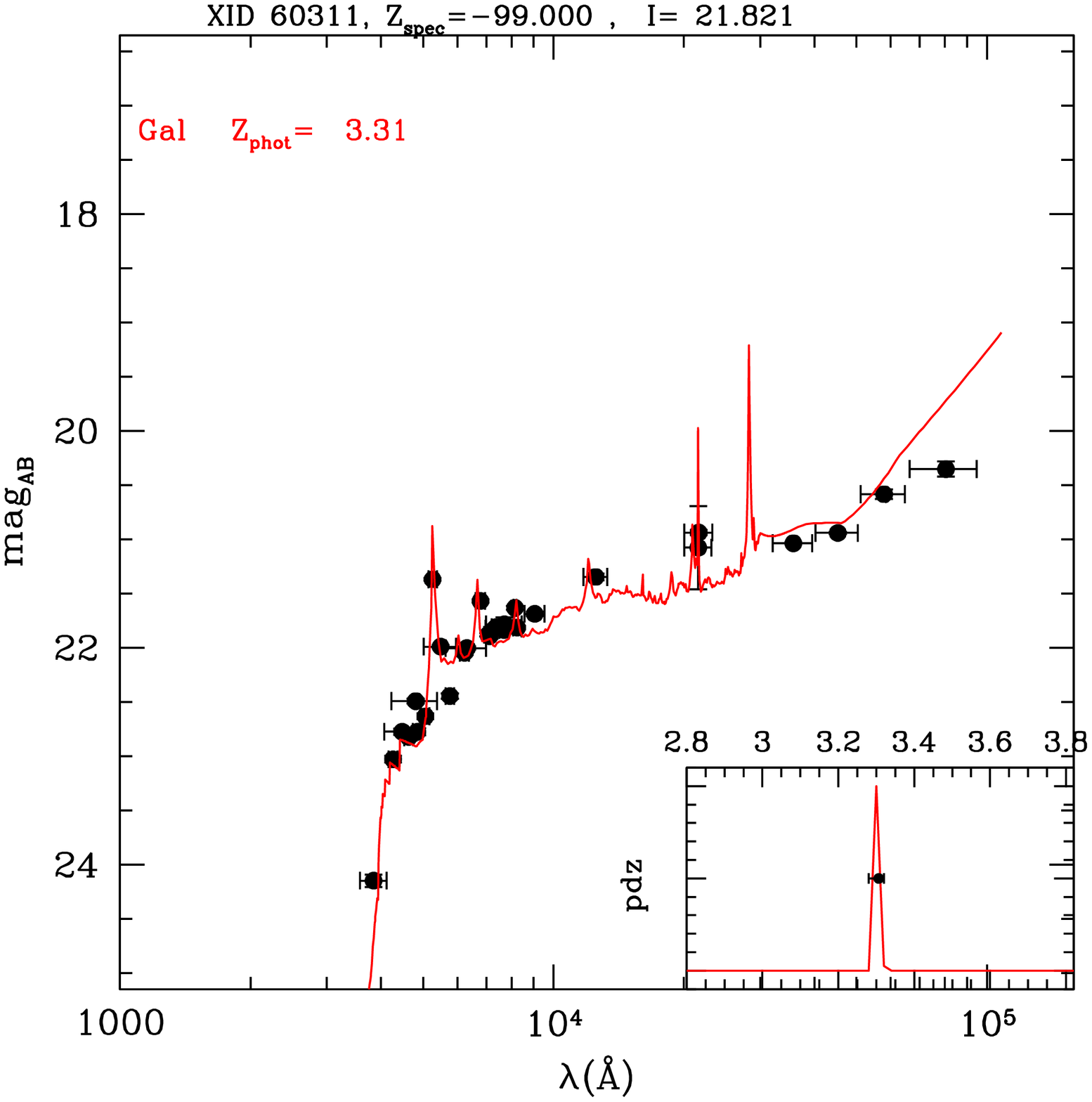}
\includegraphics[width=8cm]{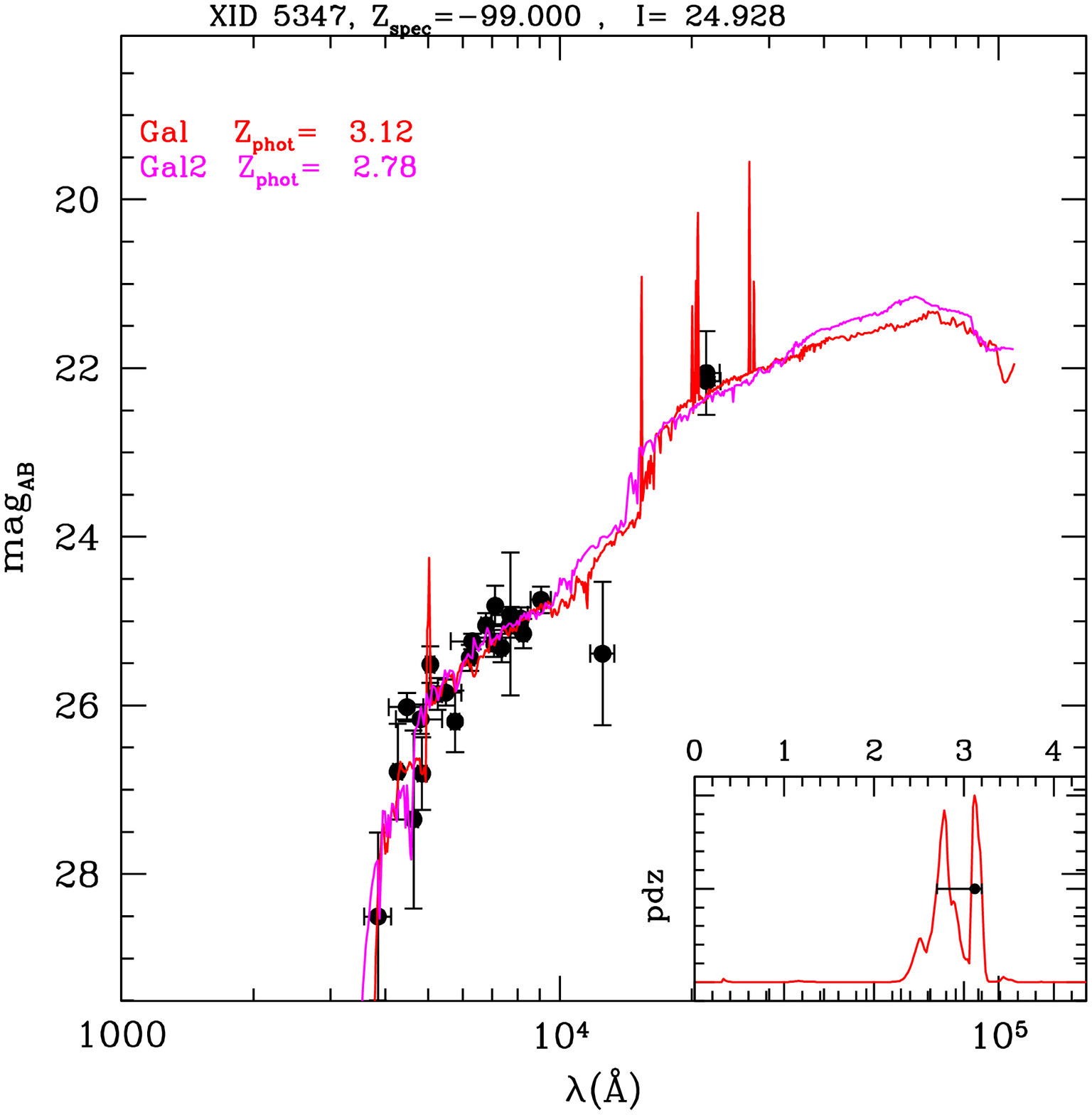}
\caption{Two examples of photometric redshift determination for sources
  without a spectroscopic redshift: a photoz with a unique solution (XID 60311)
  and an object with two possible solutions for the photometric redshift (XID
  5347). Black points are the observed photometry
  from the U-band to IRAC 8.0$\mu$m. The red curve is the best fit
  template. The inset in the bottom right shows the probability distribution
  function (normalized to one) as a function of redshift. }    
\end{figure}

\begin{figure}[!t]
\includegraphics[width=13.cm]{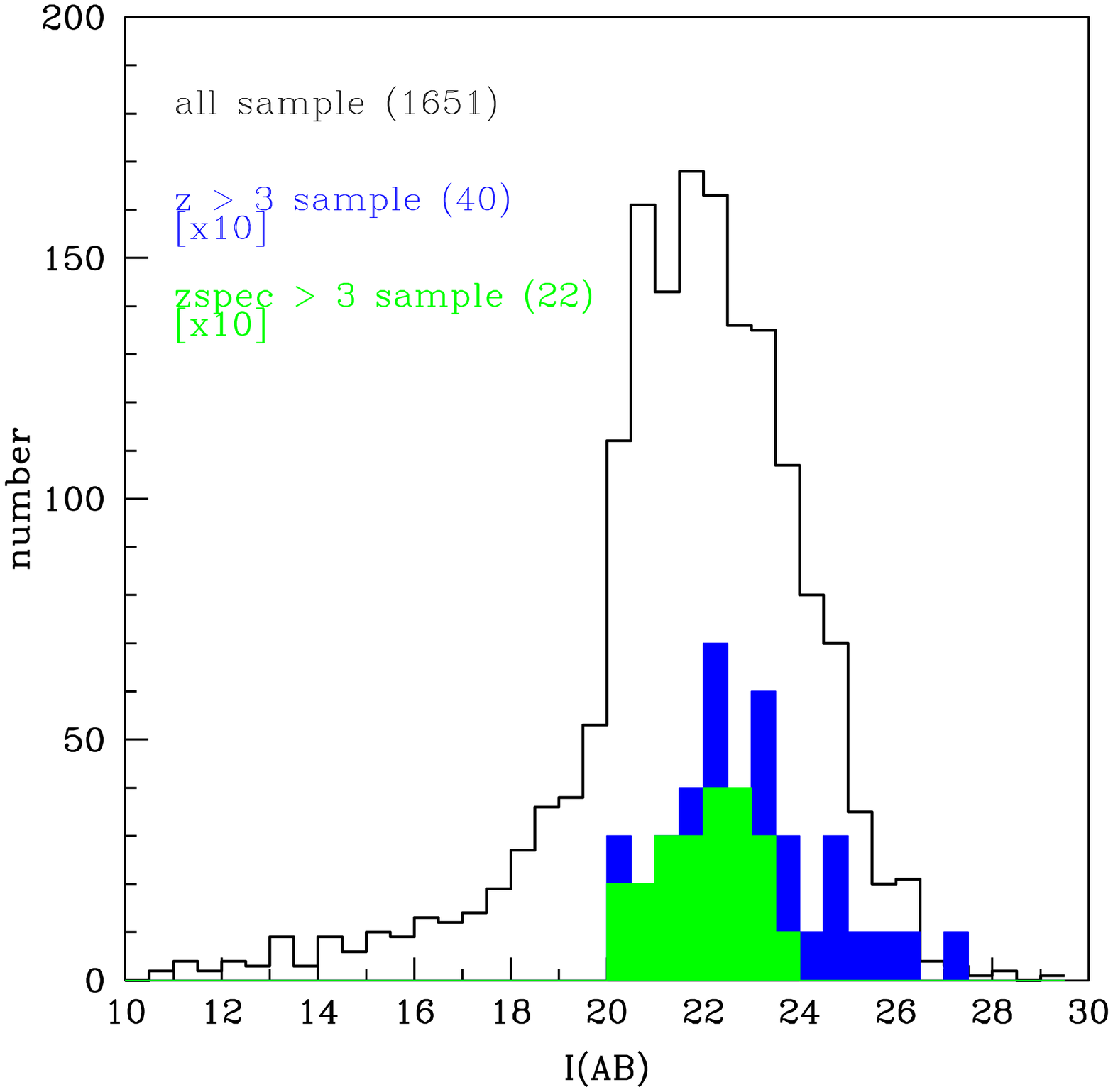}
\caption{
Histogram of the i-band magnitude distribution for the global XMM--COSMOS sample
(empty), the $z>3$ sample (blue) and the spectroscopic $z >$ 3 sample (green)}
\end{figure}
%

\subsection{The z$>3$ QSOs sample} 

We used the combined spectroscopic and photometric information available on
the redshifts in our XMM--COSMOS sample to select the sample of $z >3$ quasars. 
There are 40 XMM--COSMOS sources which have a spectroscopic (22) or, when
spectroscopic redshifts are not available, 
photometric (18) redshifts larger than $z>3$. Nineteen of them  are detected in 
both the soft and hard bands, 20 are detected only in the soft band at 
fluxes larger than 1$\times 10^{-15}$ \cgs\ and 1 object
is detected only in the hard band above the chosen threshold of
6$\times10^{-15}$ \cgs, while it is just below the adopted limiting flux in the
soft band.

Table 1 lists the name of the objects (following the standard IAU
  notation), the X--ray identifier number from the XMM-COSMOS catalog 
(Cappelluti et al. 2008) used as reference identifier in the following, 
the identifier number from the COSMOS photometric
catalog (Capak et al. 2007) and the coordinates of the optical counterparts.

The basic properties (photometric and spectroscopic redshifts, I--band
  magnitude, X--ray
  fluxes and luminosities, and hardness ratio) for these objects are reported in Table 2.
 The optical photometry used to derive the photometric redshifts
can be retrieved by the public COSMOS photometric catalog
available at IRSA\footnote{http://irsa.ipac.caltech.edu/data/COSMOS/tables/cosmos\_phot\_20060103.tbl.gz}
via the optical identifier number provided in the 3rd column of Table 1.
The deep {\it Chandra} survey in the COSMOS field covers about 
0.9 deg$^2$ in the central region. {\it Chandra} 
images are available for about half (19) of the high--z QSOs in our sample and 
for these objects the identification of the counterparts is secure, thanks to the 
positional accuracy provided by {\it Chandra} spatial resolution. 
Following the discussion in Brusa et al. (2008), we expect that, among the 21
sources without {\it Chandra} coverage, at most one can be a
misidentification. 

Spectroscopic redshifts are available for 22 sources, and are reported in
Table 2. The observed frame spectra,
sorted in order of ascending redshift, are shown in Figure 1, with all the
major emission lines labeled. 
For 4 of them, the spectroscopic redshift 
is based on a single, broad line identified as CIV1549
or Ly$\alpha$. 
In all four cases, the photometric redshift rules out 
alternative solutions at lower-redshift, favoring the proposed $z>3$
nature.
 
In all but two cases there is convincing evidence for
the  presence of broad optical lines.  
For two sources (XID 5162 and XID 5606) the most prominent emission 
lines are narrow (FWHM$<1500$ km s$^{-1}$) and thus these objects are
candidate Type 2 QSOs.   

For the remaining 18 sources, only photometric redshifts are available. 
The photometric redshifts for the entire $z>3$ sample, with associated 
errors (1$\sigma$) are reported in Table 2. 
A detailed discussion on the quality and reliability of AGN photometric
redshifts is presented elsewhere (Salvato et al. 2008). 
Here we limit to note that all but one (XID 2407) of the objects with
spectroscopic redshifts have photometric redshifts well consistent with the
spectroscopic  
ones, with ~60\% of the spectroscopic redshifts being within the 1$\sigma$ 
range of the photometric redshifts. Moreover, also the fraction of 
outliers (1 source out of 22) is consistent with
the  expectations for the entire photo-z catalog (5\%, see Section 2.1). 
However, we should also note that spectroscopic redshifts
are available for the brightest (i$\lsimeq23$) sample and 
a higher fraction of outliers (up to ~15\%) is 
expected at fainter optical magnitudes.

Salvato et al. (2008) also computed probability distribution functions 
(PDF) for the photometric redshift solution. 
In the sample of 40 quasars, five objects have the first solution (i.e. 
the highest value of PDF) at $z>3$, but a comparable solution (in terms of
PDF) at $z < 3$ exists.  
Most of these objects have i$\gsimeq24$.   
Conversely, there are additional 14 objects in the entire sample of XMM--COSMOS
counterparts which present the first solution at $z<3$, 
but have a second solution at $z>3$.   
Figure 2 shows two examples of the SED fitting and photometric redshift
determination for a source with a reliable 
solution (XID 60311, left panel) and  for an object with multiple, secondary
solutions at lower redshift (XID 5347, right panel).
Weighting the objects with the relative PDF, we obtain that 
$\sim 60$\% of the sample of the 32 objects without spectroscopic 
redshifts and with either unique,
first or second solution at $z>3$ are expected to be at $z>3$. This translates 
in $\sim20$ objects, close to the 18 objects with primary solution at
$z>3$, which are listed in Table 2. 

Among the 22 sources lacking a photometric redshift, due to problems related to
blends or saturation, six have spectroscopic redshifts which place them at
$z<1.5$. We expect that the redshift distribution of these sources follows the
one of the total sample, with only 3\% (22/683, e.g. $\sim1$ object) being at
$z>3$. It should also be noted that the 14 objects undetected in the optical
bands (10 of them detected in the soft band above the adopted limiting flux)
and without an estimate of the photometric redshift, are candidate high
redshift, possible obscured QSOs (see, e.g. Koekemoer et al. 2004, Mignoli et
al. 2004).   

Summarizing, we conclude that, although the fraction of objects in our 
sample which have only a photometric redshift is close to 50\%, the total 
number of objects with $z>3$ is statistically robust. Even if some of the 
18 objects without spectroscopic redshift are likely to be at $z<3$, 
this number is expected to be almost exactly compensated by the number 
of objects which, being at $z>3$, have instead their best photometric 
solution at $z<3$. Therefore,  we consider the 40 objects listed in 
Table 2 as representative of the real number of sources at $z>3$. We 
will quantify in Section 6 the possible contribution 
  of objects undetected in the optical bands to the total number of $z>3$
  QSOs.


Figure 3 shows the magnitude distribution of the $z>3$ sample (blue filled
histogram) compared with the overall optical population (black open
histogram). 
The median magnitude of the $z>3$ sample (i=22.72, with a dispersion of
1.02 mag) is about 0.8 magnitudes fainter than that of the overall XMM--COSMOS
population ($i=21.90$, with a dispersion of 1.32). 
Spectroscopically confirmed $z >$ 3 objects (green histogram in the figure) are in the
bright tail of the magnitude distribution of the high redshift sample. 
A non negligible fraction of the objects in the sample (7/40) has
$i>$ 24.5 and, as pointed out above, a somewhat more uncertain photo-z solution
(see Table 2).   

\section{Optical colors} 

Multiband optical photometry 
provides a routinely-adopted, reliable selection technique 
to select high redshift QSOs. 
The method was first applied to AGN in the
1960s, based on the inference that quasars often have a larger ultraviolet
excess than the hottest stars (Sandage \& Wyndham 1965), and subsequently 
extended to large-scale surveys (e.g. Schmidt \& Green 1983, Croom et al. 2001,
Fan et al. 2001, Richards et al. 2006). 
The most suitable combination of colors strongly depends on the redshift range.
For example, the  "UV excess" technique is optimal for $z<$ 2.0--2.2.
At  $z > $ 3 an efficient selection method is known as Lyman Break technique
and has been extensively used to select $z >$ 3 QSOs and galaxies (see, e.g.,
Steidel et al. 1996, Hunt et al. 2004, Aird et al. 2008).
Assuming a standard QSO template and including absorption by the Intergalactic Medium 
(IGM, i.e. the Lyman-alpha forest), it is possible to efficiently  isolate $z >$ 3  QSOs 
in the $U-g^{\prime}$ vs $g^{\prime}-r^{\prime}$ color-color plane
(i.e. Siana et al. 2008). 

A somewhat different color--color plot ($v-i$ vs. $B-v$) has been extensively
discussed  by Casey et al. (2008): 
for efficient selection at $z>3$ they make a diagonal cut in the two color
diagram, described by the relation: $v-i<1.15\times(B-v)-0.31$. 
The three photometric bands  have been chosen 
in order to minimize the problems due to AGN variability: COSMOS observations 
in the $B,v$ and $i$ bands were taken at the same epoch, providing
almost simultaneous colors, while the $g'$ and $U$ band observations were
taken about one year apart (Taniguchi et al. 2007, Capak et al. 2007). 

\begin{figure}[!t]
\includegraphics[width=8cm]{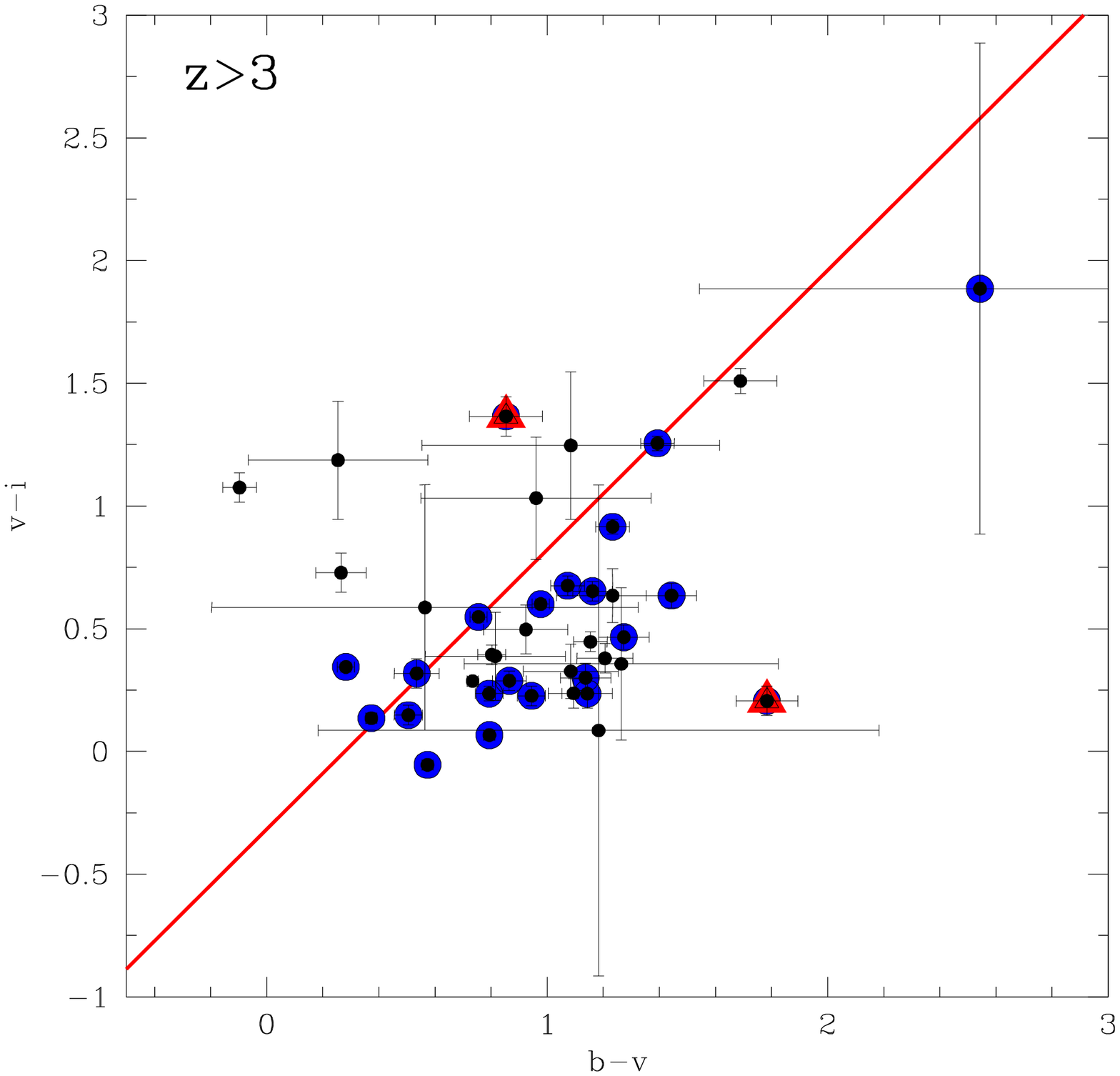}
\includegraphics[width=8cm]{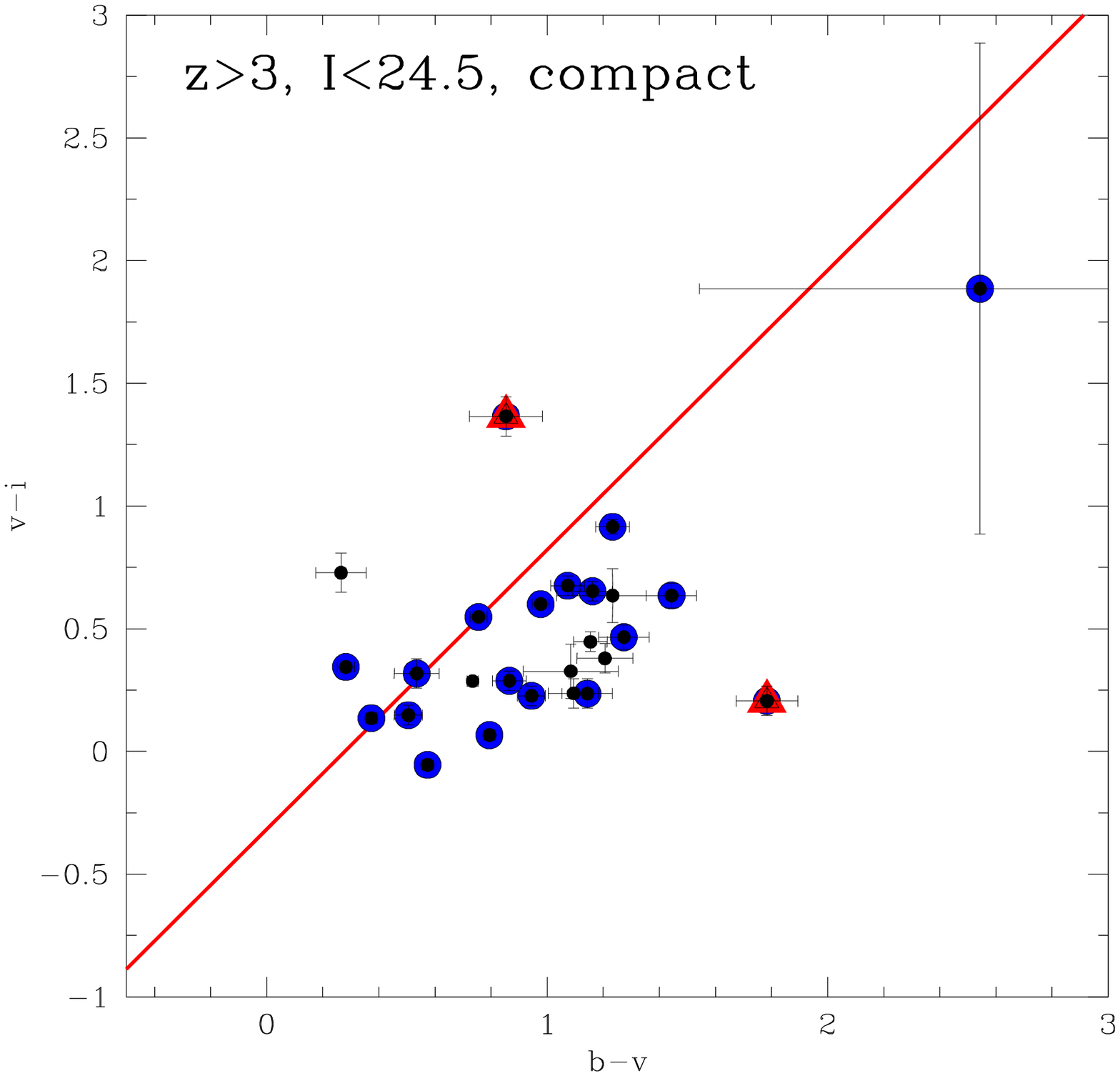}
\caption{
{\it Left panel}: $v-i$ vs. $B-v$ color--color diagram for the $z>3$ sources. 
Black points correspond to all $z>3$ sources, blue circles to the
20 objects classified as BL AGN, red triangles refer to the
2 objects classified as NL AGN. 
Outliers from the QSOs locus are those on the left of the 
red solid line, defined as:  [$v-i<1.15\times(B-v)-0.31$]. {\it Right panel}:
same as the previous panel, but only for the subsample with $i<24.5$ and
compact morphology.}  
\end{figure}
%

The X--ray selected QSOs of our sample are plotted in
the $v-i$ vs. $B-v$ plane in Figure 4 (left panel) with associated errors.
The 20 spectroscopically confirmed BL AGN are marked with blue circles,
while the 2 objects classified as NL AGN are shown as red triangles.  
Eight objects ($\sim 20$\%) lie above the nominal line proposed by Casey et al. (2008)
to select $z>3$ QSOs in the COSMOS field. 
Among them two sources (XID 2407 and XID 5606) have a spectroscopic redshift.
These objects would have not been selected as high
redshift quasars from this optical color--color diagram. 
We should note, however, that the Casey et al. (2008) criterion 
was tested for relatively bright ($i<$ 24.5) sources with compact morphology
defined in terms of Gini coefficient larger than $\sim0.8$.
If we further impose the same optical limit to the objects of our sample, and
consider only the objects classified as point-like from the ACS analysis
(Leauthaud et al. 2007), only 3 out of 25 ($\sim$ 12\%) lie above the dividing
line (Fig.~4, right panel).  
The eight objects above the dividing line in the left panel of Fig.~4 are, 
therefore, on average, optically faint ($i>$ 24.5) and/or with an extended
morphology. These objects would not have been selected as high redshift candidates 
by an optical survey.
We name these objects ``outliers'' and investigate their average X--ray properties 
in the following. 
Conversely, if we consider the full sample of  XMM sources 
with  $i<24.5$ and  ``compact'' morphology, there is a total of 60 
objects that lie below the diagonal line in Fig.~4:
25 shown in the right panel
of Fig.~4 plus 35 additional objects. The majority of these 35 objects (25/35) 
has a good quality spectroscopic redshift lower than 3, and in most of the cases
(20/25) they are classified as BL AGN at $z \sim$ 1--3.
As discussed by Casey et al. (2008) the contamination of low--redshift sources
in this optical color--color diagram can be as high as 50--70\%.

\section{X--ray properties}

For each object with given redshift (either spectroscopic or
photometric) the flux in the rest frame 2--10 keV band has been
calculated, assuming the spectral slope derived from the ratio of the 
fluxes in the rest-frame hard and soft band\footnote{Within   each 
observed band the counts are converted into fluxes assuming $\Gamma=2$ in the
0.5-2 keV and $\Gamma=1.7$ in the 2-10 keV, as in Cappelluti et al. 2007.}.  
Rest frame 2--10 keV fluxes have been then converted into luminosities
within the adopted cosmology. Figure 5 (left panel) shows the histogram of the
rest frame, 2--10 keV luminosities for the 40 sources in the present sample.  
The rest frame 2--10 keV luminosities calculated using the observed 
0.5--2 keV fluxes (roughly corresponding to the rest frame 2--10 keV fluxes at z$\geq 3$)
are consistent with those derived with the method described above
(see black histogram in Figure 5). 
All the objects in the $z > 3$ sample have hard X--ray (2--10 keV) rest--frame 
luminosities in the interval 10$^{44}$-10$^{45}$ erg s$^{-1}$. 
The luminosity-redshift plane in the redshift interval $z$=3.0--4.5
is shown in Figure 5 (right panel).
Sources with a spectroscopic redshift are plotted with blue (20 BL AGN)
and red (2 NL AGN) symbols. 
The continuous line represents the luminosity limit of the survey 
computed from the 0.5--2 keV limiting flux.

The Hardness Ratio HR (defined as (H--S)/(H+S), where H and S are the counts
detected in the 2--10 keV and 0.5--2 keV bands, respectively) as a function of
the redshift for all the 40 objects in our sample is shown in Figure~6. 
For the 20 sources detected only in the soft band, upper limits have been 
obtained by conservatively assuming 25 hard counts.
Loci of constant $N_H$ at different redshifts 
for a power law spectrum with $\Gamma=1.7$
are also reported with dashed lines\footnote{We did not consider a reflection
  component because it is not observed at the X--ray luminosities considered in this work.}.
The shaded region marks the expected HR for an unabsorbed power-law with
$\Gamma=1.7$ (upper limit) and $\Gamma=1.8$ (lower limit).
The objects in our sample span a rather large range of HR and corresponding
$N_H$ values, from unobscured to column densities up to logN$_{H} \simeq 23.7$. 

To increase the statistics and gain information on the average spectral
properties of the different sources in the sample, we performed a stacking
analysis of the source spectra. The spectra were extracted for all the sources
in their rest-frame 2--22 keV band from the EPIC {\it pn} data and corrected for the
local background, following the procedure outlined in Mainieri et al. (2007). 
Three sources (XID 5259, 5592, and 5594),
which have null or negative counts in this energy range when the local
background is subtracted, are excluded from the analysis here. 
Two additional sources (60186 and 60311) which do not show stringent upper
  limits in the HR are further excluded. The individual
spectra were then corrected for the instrumental response curve, brought to
the rest-frame and stacked together with energy intervals of 1 keV. Noisy
spectral channels of the stacked spectra were binned further, when
appropriate. 

\begin{figure}[!t]
\includegraphics[width=8.cm]{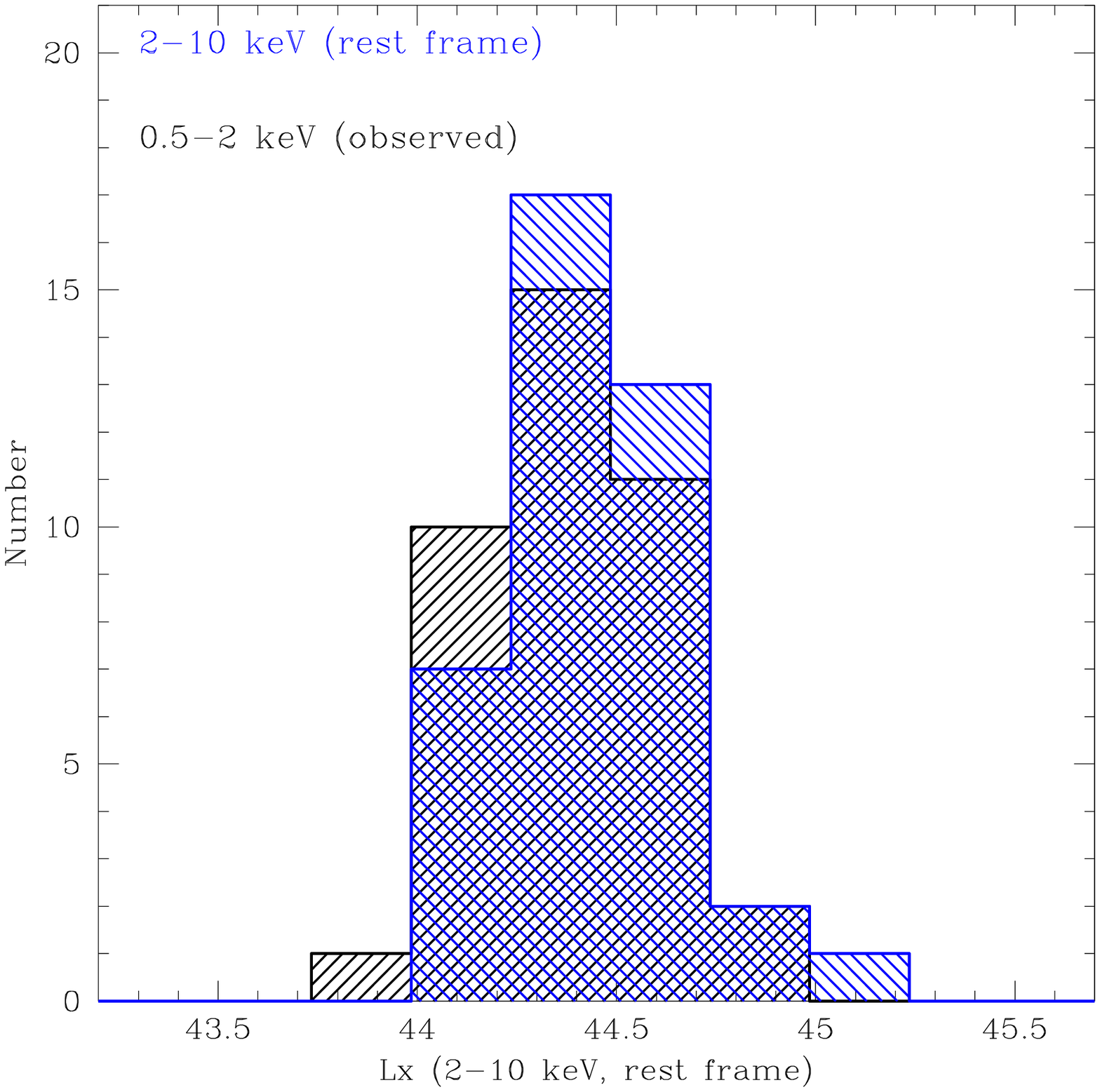}
\includegraphics[width=8.cm]{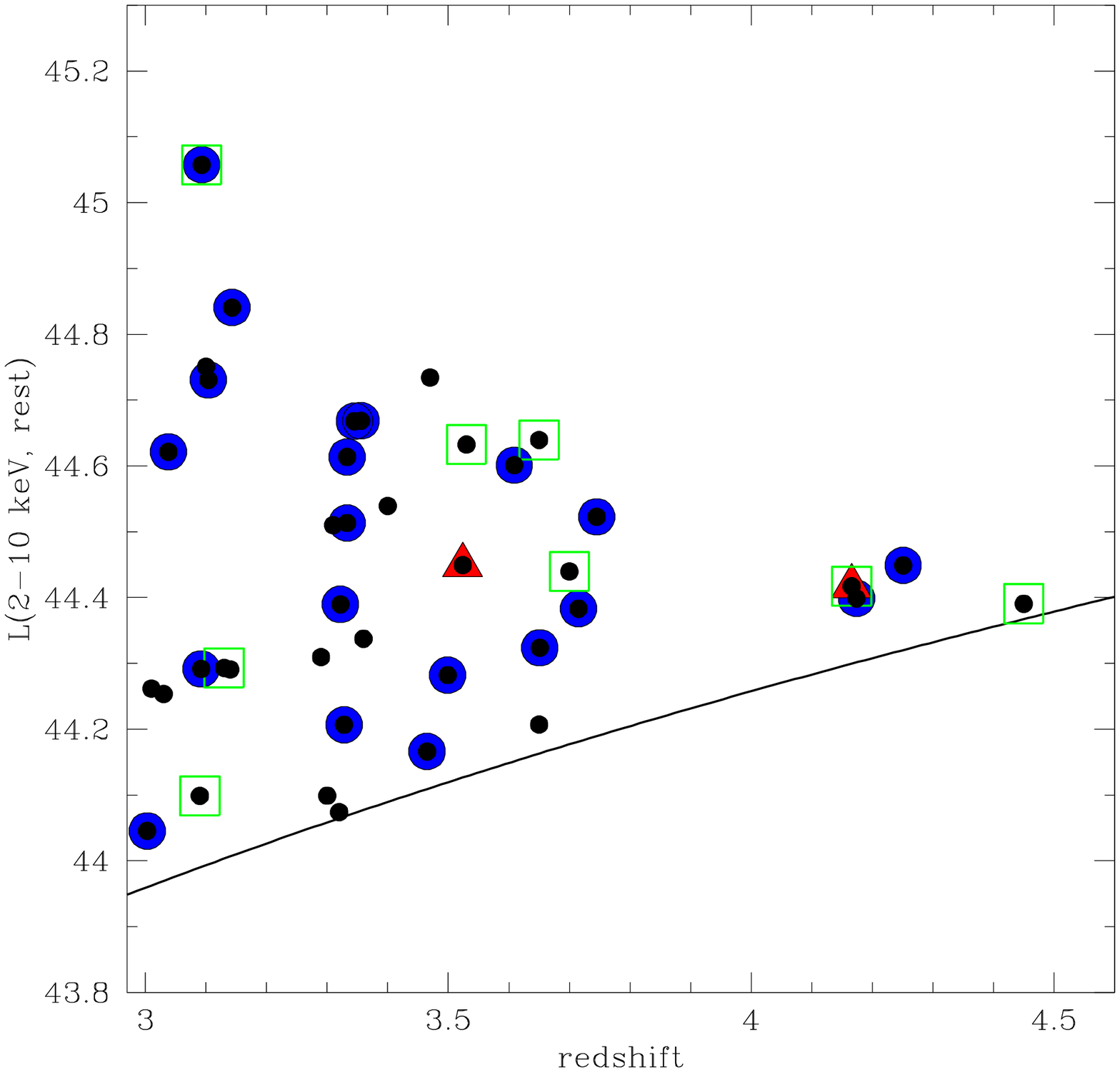}
\caption{
{\it Left panel}: Histogram of the rest frame 2--10 keV luminosity 
calculated assuming the spectral slope derived from the ratio 
of the fluxes in the rest frame soft and hard bands (blue histogram, see text
for details) compared with the rest frame 2--10 keV luminosities calculated
using the observed 0.5-2 keV fluxes (i.e., the rest frame 2-10 keV flux at
z$>3$, black histogram). {\it Right panel}: the luminosity redshift plane
for the objects in our sample. Symbols as in previous figures. Green squares
mark the outliers from the optical color-color diagram of Fig.~4.
The continuous line represents the luminosity limit of the survey 
computed from the 0.5--2 keV limiting flux. 
} 
\end{figure}

\begin{figure}[!t]
\includegraphics[width=13.cm]{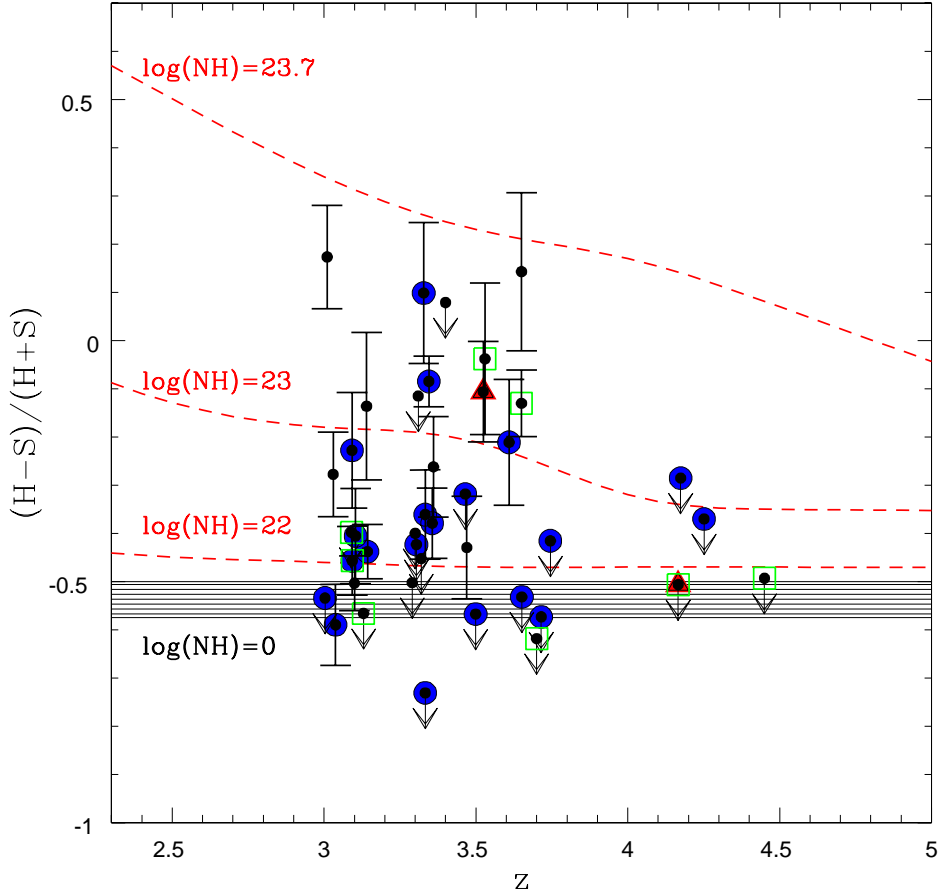}
\caption{
HR vs. redshift for the $z > 3$ QSOs sample (symbols as in previous
figures). The dashed lines correspond to the loci of constant $N_{\rm H}$ 
for a power-law spectrum with $\Gamma=1.7$. The shaded region marks the
expected HR for an unabsorbed power law with $\Gamma=1.8$ (lower limit) and
$\Gamma=1.7$ (upper limit). }
\end{figure}
%

At first, we divided the sample into two sub-samples based on the HR: the
sources showing an indication of ``soft" spectrum from the HR (i.e., the 26
objects below the locus of $N_{\rm H} = 10^{23}$ cm$^{-2}$ at different
redshifts) and the sources with an indication of a ``hard" spectrum (i.e., the
9 objects above the $N_{\rm H} = 10^{23}$ cm$^{-2}$ locus). The stacked
spectra for these two sub--samples are shown in Figure~7 and the results from a
power-law fit are summarized in Table~3. The ``soft" sources (open squares in
Fig.~7) have a steeper spectral slope ($\Gamma\sim 1.8\pm0.3$) than the ``hard"
sources (filled squares: $\Gamma\sim 0.9\pm0.8$), demonstrating that the HR analysis
(which is based on the X--ray color in the observed frame) can isolate the
most absorbed sources even at these high redshifts. 

The fraction of log $N_{\rm H}\geq 23$ AGN, computed weighting with the
errors the number of sources with HR above the log$N_{\rm H}$=23, is  
$\sim 20$\%, somewhat larger than 
the Gilli et al. (2007) expectations\footnote{The expectations have been
  computed using the Portable Multi Purpose Application for the AGN COUNTS
  (POMPA-COUNTS) software available on-line at the link:
  http://www.bo.astro.it/~gilli/counts.html.}  
 ($\sim10$\%) for the same luminosity, redshifts and flux limits of our survey. 

It is interesting to note that one of the two objects with narrow emission
lines in the optical spectrum (XID 5162, z=3.524) has an HR ($\sim -0.11$, red triangle 
in Fig.~6) consistent with a
column density log$N_H >23$  and therefore
would be classified as a Type 2 QSO from both the optical and the X-ray
diagnostics. This object is one of the highest 
redshift, radio quiet, spectroscopically confirmed Type 2 QSO (see e.g.
Norman et al. 2002, Mainieri et al. 2005).  
The other two objects with log$N_H>$ 23 and spectroscopic redshift
(XID 60131 at z=3.328, and XID 1151 at z=3.345) show clear signatures of broad
  absorption lines (BAL) in their optical spectra (Fig.~1). Column densities
  of the order of log$N_H \sim$ 23 are common among those BAL QSOs for which
  good quality X--ray spectra are available (Gallagher et
  al. 2002). Significant X--ray absorption in BAL QSOs is detected up to $z
  \sim$ 3   (Shemmer et al. 2005).   

It is important to note, however, that the HR at $z > 3$ is not very sensitive
to column densities below log$N_{\rm H}\simeq$23.  
Given also the typical dispersion in the intrinsic power-law spectra, it is
possible that the sub-sample of ``soft'' sources contains also some of these
absorbed  AGN. These sources are expected to be optically faint 
(e.g. Alexander et al. 2001, Fiore et al. 2003, Mainieri et al. 2005, Cocchia
et al. 2007) and thus among the still spectroscopically 
unidentified sources. In order to test this possibility, we further divided
the ``soft'' sample into two subsamples: sources without a spectroscopic
redshift available (10) and sources spectroscopically identified as BL AGN 
(16, see Figure 7, right panel). 
There is evidence of absorption (below $\sim$ 3 keV, corresponding to
log$N_{\rm H}\sim$22.5) in the average spectrum of spectroscopically
unidentified sources. 
The X--ray counting statistics does not allow to establish, on a source by
source basis, whether the "outliers" of optical color--color diagrams are also
obscured in the X--rays. 
Given that the five "outliers" from the optical color--color diagram make up
about 30\% of this absorbed sample, it may well be possible that they are responsible 
of the low energy cut--off in the stacked spectrum (right panel of Fig.~7).

\begin{figure}[!t]
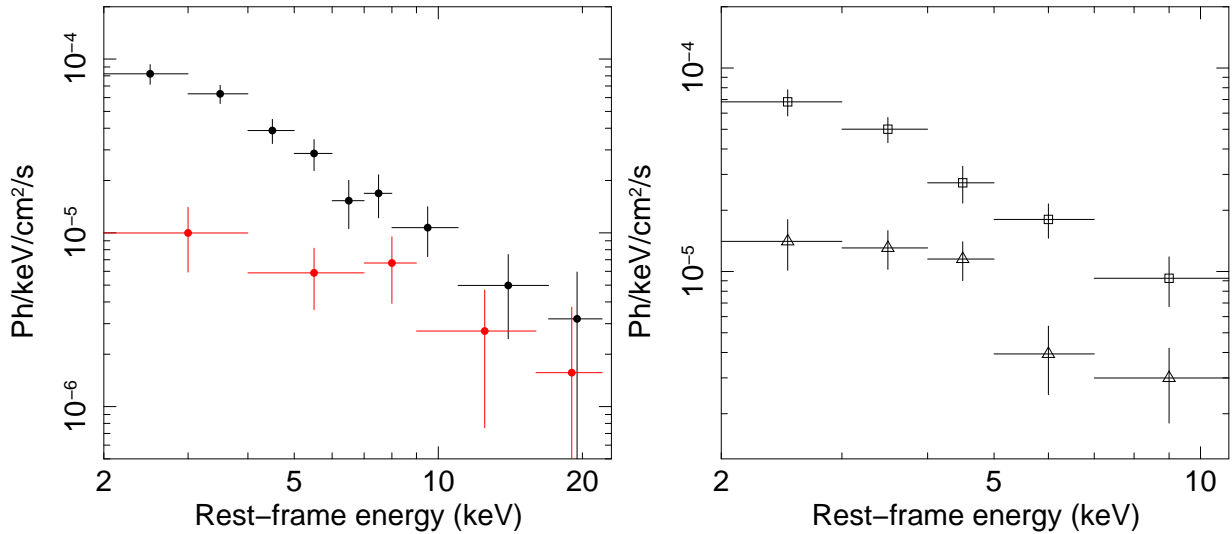

\includegraphics[width=7.cm,angle=-90]{f7a.eps}
\includegraphics[width=7.cm,angle=-90]{f7b.eps}
\caption{
{\it Left panel}: Rest-frame stacked spectra of sources with logNH$<23$ at all
redshifts (26 objects, black circles) and with logNH$>23$ (9 objects, red
circles). {\it Right panel}: Rest-frame stacked spectra of the sources with
with logNH$<23$ further divided in BL AGN (16 objects, squares) and
spectroscopically unidentified sources (10 objects, triangles).}
\end{figure}
%

\section{Evolution of $z>3$ QSOs}

\subsection{Number counts}

We derived the logN--logS of the $z>3$ XMM--COSMOS QSOs 
by folding the observed flux distribution with the sky coverage.
The binned logN--logS relation of the 39 objects with a 0.5--2 keV flux larger than 
10$^{-15}$ erg cm$^{-2}$ s$^{-1}$ is plotted in Figure~8 (red points, with associated
Poissonian errors). 
The green shaded area represents an estimate of the maximum and minimum number
counts relation at $z >$ 3 under somewhat extreme assumptions. 
The lower bound is obtained by considering only the 22 sources 
with a spectroscopic redshift\footnote{We included also 
the four objects with a ``single line''spectrum, given that the photometric 
redshift solution discarded alternative solutions at lower-redshift}, 
while the upper bound
includes in the $z>$ 3 sample also the 10 objects with a 0.5--2 keV flux larger than
10$^{-15}$ \cgs and without a detection in the optical band and the
11 objects above the same flux threshold  and with a second photometric
redshift solution at $z>3$ (see $\S$2). 
Under these assumptions, the lower limit corresponds to the (very unlikely)
hypothesis that all the photometric redshifts are overestimated, while the
upper limit corresponds to the assumption that the non--detection in the
optical band is a very reliable proxy of high redshift in X-ray selected
samples  (e.g. Koekemoer et al. 2004; Koekemoer et
al. 2008).    
The dashed blue curve corresponds to the predictions
of the XRB synthesis model (Gilli et al. 2007), obtained
extrapolating to high-z the best fit parameters of Hasinger et al. (2005),
while the solid blue curve represents the model predictions obtained 
introducing in the LF an exponential decay with the same functional form
($\Phi(z) = \Phi(z_0) \times 10^{-0.43(z-z_0)}$) adopted by Schmidt et
al. (1995) to fit the optical LF, corresponding to an e--folding per unit 
redshift. 

The model predictions at $z >$ 3 obtained extrapolating the Hasinger et
al. (2005) best fit parameters at high-z (dashed line) clearly overestimate
the observed counts even in the  most optimistic scenario 
(upper boundary in Fig. 8). 
A much better description to the observed counts is obtained by introducing
the exponential decline in the X--ray LF at $z >$ 2.7 (i.e. we have here
adopted $z_0=2.7$ following Schmidt et al. 1995). 
 
An estimate of the number counts for $z > $4 QSOs is also reported in Fig.~8. 
Although only 4 objects are detected in the XMM--COSMOS survey, there is a 
remarkably good agreement with the predicted number counts, suggesting that
the Schmidt et al. (1995) parameterization provides a good description of the  
X--ray selected QSOs surface densities up to $z \simeq$ 4.5. 

\begin{figure}[!t]
\includegraphics[width=13.cm]{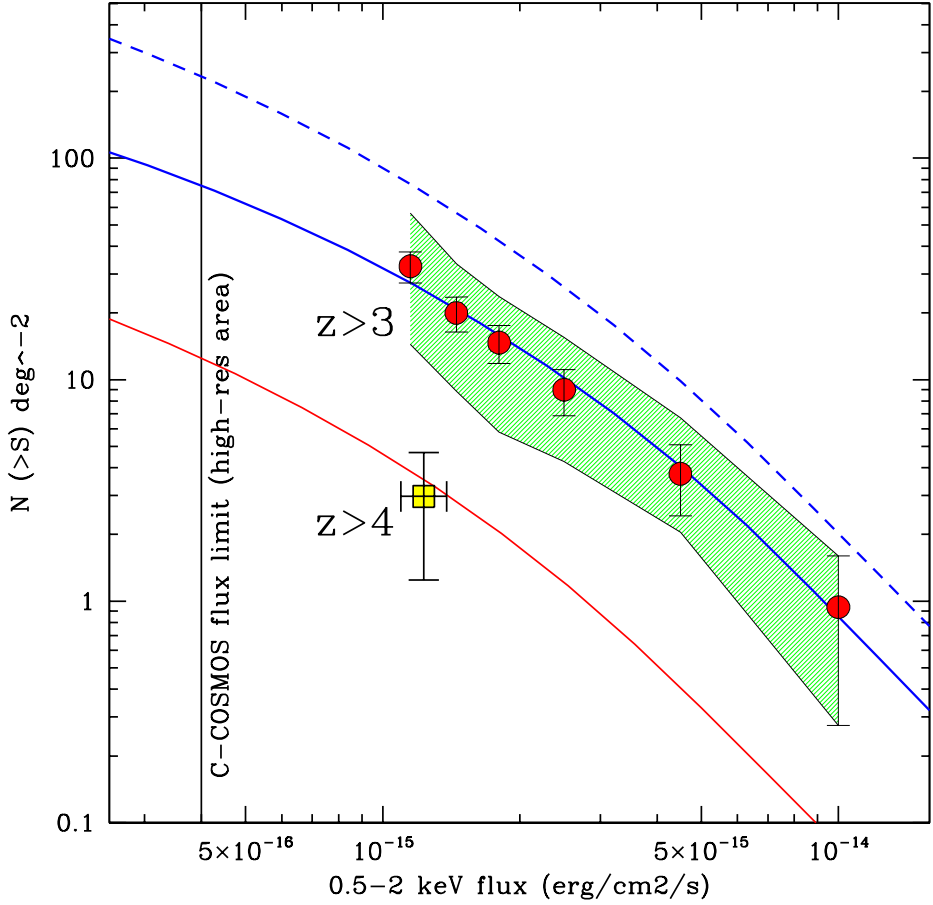}
\caption{The binned LogN-logS relation (red circles with associated
Poissonian errors) of the z$>3$ QSOs population. 
The dashed and solid blue curves correspond 
to two different predictions based on the Gilli et al. (2007)
model. 
The former is obtained extrapolating the best fit parameters of Hasinger et
al. (2005) for the evolution of the X--ray LF to high-z, while  
an exponential decay at $z >$ 2.7 is introduced in the latter.
The green shaded area represents an estimate of the global error budget (see
text for details). 
The yellow square (with associated Poissonian error) is the estimate of the 
number counts for $z >$ 4 quasars based on the 4 objects (3 of them with 
a spectroscopic redshift) in the XMM--COSMOS
sample. The red curve shows the predicted number counts assuming 
the exponential decay in the LF. The vertical line at $4\times10^{-16}$ \cgs\
represents the limit of the C-COSMOS survey.}
\end{figure}
 
\subsection{Comoving space densities}

The comoving space density of $z>$ 3 QSOs in three, almost equally populated, 
redshift bins ($z = 3.0-4.5$) is reported in Figure 9. 
The comoving space densities have been computed with the 
1$/V_{max}$ method, originally proposed by Schmidt
(1968), which takes into
account the fact that more luminous objects are detectable over a larger
volume.  
In order to reduce the effects of incompleteness (see Fig.~5 right panel),
the space densities have been computed for luminosities logL$_X > 44.2$ 
(i.e. considering only 34 out of 39 objects).  
Because of the significant effects introduced by X-ray absorption, some 
particular care is needed in computing V$_{max}$ for X-ray selected sources. 
In fact, for any given intrinsic luminosity, the effective limiting flux for
detecting obscured sources is higher 
than for unobscured ones and the corresponding volume over which 
obscured sources can be detected is smaller. 
We therefore calculate the maximum available volume for each source using 
the formula:

\begin{displaymath}
V_{max}=\int_{z_{min}}^{z_{max}} \Omega(f(L_x,z,N_{H}))\frac{dV}{dz}dz
\end{displaymath}

where $\Omega(f(L_x,z))$ is the sky coverage at the flux $f(L_x,z)$
corresponding  to a source with absorption column density $N_H$ and 
unabsorbed luminosity $L_X$, and $z_{max}$ is the maximum redshift at which
the source  can be 
observed at the flux limit of the survey\footnote{If $z_{max}>z_{up,bin}$,
  then $z_{max}=z_{up,bin}$}. 
More specifically, for unabsorbed sources we adopted the observed rest--frame
2--10 keV  luminosity (see $\S$4), while for obscured ones the unabsorbed
luminosity was derived  assuming the best fit column density as obtained from
the hardness ratio. 
The limiting flux for each absorbed source was computed by folding the
observed spectrum with the XMM sensitivity. 

Secondly, for a proper comparison with model predictions at a given
luminosity, we should consider the population of obscured sources that are
pushed below the limiting flux  by the X--ray absorption, and correct the
observed space density accordingly. 
In order to estimate the number of missed sources we considered a population
of luminous (intrinsic log$L_x>44.2$) obscured QSOs with the luminosity
function and the column density distribution assumed by Gilli et al. (2007), 
and folded this population with the survey sky coverage. The results indicate
that, at 0.5-2 keV fluxes larger than 10$^{-15}$ \cgs, about 
25\% and 45\% of the full AGN population with intrinsic logL$_X>44.2$ are
missing from our sample at a redshift of 3 and 4, respectively. 
For the computation of the space density we therefore weighted 
the contribution of each object taking into account the incompleteness
towards the most obscured sources as a function of both redshift and 
source flux. 

The space density of quasars with log$L_X>44.2$ from the full XMM--COSMOS $z>3$
sample, corrected for the incompleteness against the 
most obscured sources as described above, is compared with the predictions 
from the Gilli et al. (2007) model 
at the same luminosity threshold (red dashed curve).  Obscured,
Compton thin AGN are included following the prescriptions described above;
Compton Thick AGN have not been included in the model given the 
fact that X--ray selection at the XMM--COSMOS limiting flux 
is insensitive to this population (see e.g. Tozzi et al. 2006). 
The red continuous line 
represents the expectations of the same model when the
exponential decline in the X--ray LF discussed in $\S$5.1 is introduced. 

In agreement with the results obtained from the logN--logS, the declining
space density provides an excellent representation of the observed data. 
Including the optically undetected sources in the $z>3$ QSOs 
sample, the space density of high z QSOs remains significantly below the extrapolation of 
the best fit Hasinger et al. (2005) LF. However, the shape of the observed 
decline would be different.

\begin{figure}[!t]
\includegraphics[width=12cm]{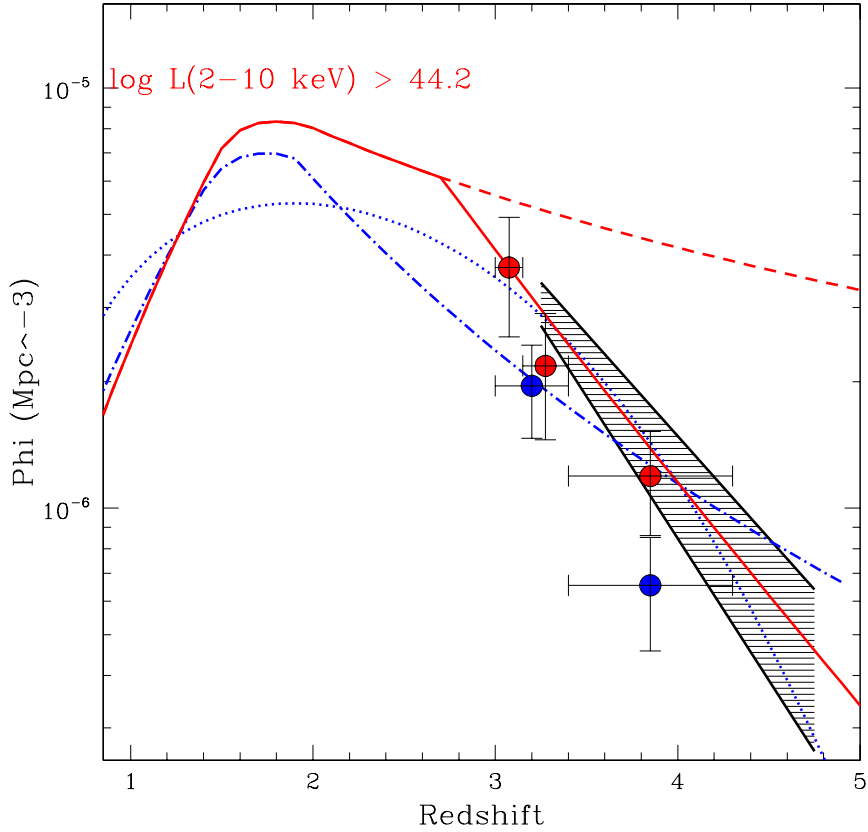}
\caption{
The comoving space density in three ($z$ = 3.0--4.5) different  
redshift bins for the $z >$ 3, log$L_X>44.2$ XMM--COSMOS sample. 
Red symbols (with associated errors) refer to the full sample
(34 objects) and have been corrected for incompleteness against obscured
sources (see text for details).      
The red curve corresponds to the X--ray selected AGN space density computed for
the same luminosity limit from the Gilli et al. (2007) model. 
At $z>2.7$ we plot with two separate curves expectations
when the exponential decline in the X--ray LF at $z >$ 2.7 is introduced
(continuous line) or not (dashed line, see text for details).  
Blue symbols (and associated errors) refer to the $i < 24$ sample 
(28 objects) and are corrected only for the volume effects to be compared with
the Silverman et al. (2008) LF. 
The dot-dashed and dotted lines correspond to the Silverman et al. (2008) 
LF, in the case of the LDDE and mod-PLE model, respectively (see Silverman et
al. 2008 for details). 
The black shaded area represents the shape derived from brighter optical quasars
surveys (Schmidt et al. 1995  and Richards et al. 2006, upper and lower bounds
respectively), rescaled up by an arbitrary factor $\sim 80$ for an easier comparison with
the X--ray data.} 
\end{figure}

In order to compare our findings with the Silverman et al. (2008) recent
estimates of the high-redshift LF, we have considered the same limit they
adopted in the 
optical magnitude ($i < 24$) without correction for the incompleteness
towards the most obscured sources. 
The blue symbols in Fig. 9 are the space densities derived from the 28 objects
at $I<24$, computed correcting only for volume.
The mod-PLE Silverman et al. (2008) LF is higher than 
our data over the entire redshift range ($3<z<4$); their LDDE LF is 
instead in very good agreement with our data point in the first redshift 
bin, while it is a factor of $\sim2$ higher than our data at $z >3.5$.

\section{Results and Discussion}


The analysis of optical colors and X--ray spectra indicates that
high redshift, X--ray selected QSOs have optical properties not
significantly different from those of optically selected, bright QSOs,
even if a non-negligible fraction ($\sim 20$\% i.e. the outliers of fig.~4) 
would not have been selected by the Casey et al. (2008) optical color-color
criteria.  
Not surprisingly, these sources are optically faint and 
make up a significant fraction (30\%) of the spectroscopically unidentified 
objects for which the observed HR does not allow to constrain the column
density. 


Even if the  bulk of the population of obscured AGN responsible for
the X--ray background is not fully sampled at the limiting flux of the
XMM--COSMOS survey, the relative fraction ($\sim$ 20\%) of obscured
(log$N_H > $ 23) QSOs is not inconsistent, given the small number statistics,
with that ($\sim$ 10 \%) predicted by Gilli et al. (2007).
We note, moreover, that the log$N_H>23$ sample might be 
contaminated by less obscured sources (log$N_H=22-23$) given
the relatively large errors associated to the measured HR. 
The depth of the XMM--COSMOS survey does not allow us to further investigate 
the absorption distribution for column densities 
lower than log$N_H \simeq$ 23, nor to firmly establish whether optically 
faint sources are X--ray obscured. 
The fraction of outliers in the optical color--color diagram is very similar 
to that of X--ray obscured sources and both are close to 20\%. 
However, the two subsamples
are marginally overlapping (only two optical outliers are also X--ray
obscured) suggesting that gas absorption and dust reddening 
are not tightly correlated (see Maiolino et al. 2001). 


The X--ray luminosity function of hard
X--ray selected AGN as determined by combining several {\it Chandra} and XMM surveys
(Ueda et al. 2003; La Franca et al. 2005)  is well constrained up to
relatively low redshifts ($z\sim$ 2--3). 
At higher redshifts, the number statistics has so far prevented 
a robust estimate of their space density. Evidence for a decline at $z >$ 3 
was reported by Hasinger et al. (2005) and Silverman et al. (2005), but 
is limited to bright, unobscured QSOs.
The $z >$ 3 QSOs sample drawn from the XMM--COSMOS survey allowed us, for the
first time, to firmly address the issue of the evolution of high redshift QSOs
thanks  to a homogeneous and sizable sample of X--ray sources  
much less biased with respect to mildly obscured AGN and with an almost 
complete redshift information. 
The results indicate that the comoving space density of X--ray luminous 
($L_X \gsimeq 10^{44}$ erg s$^{-1}$) QSOs at $z\sim 3$ is 
($3.7\pm1.2)\times10^{-6}$ Mpc$^{-3}$ and declines exponentially (by an
e--folding per unit redshift) in the $z \sim$ 3.0--4.5 range.
These results appear to be robust, despite 
the still remaining (small) uncertainties on the photometric 
redshifts discussed in Sect. 2.2. Moreover, if 
all the sources
  undetected in the optical band were at $z>3$ equally populating the
  different redshifts bins, the shape of the space density as a function of redshift    would not change
  significantly.
  Only in the case that all the 10 optically undetected, soft X--ray
  sources turn out to be at $z>$3.5-4, the analytical parameterization of the
  observed decay should be modified.

The high--redshift decline is similar to that of luminous ($M_I
< -27.6$), optically bright, unobscured QSOs as well established by SDSS 
observations (Richards et al. 2006) and can be satisfactorily described 
by the Schmidt et al. (1995) parameterization. 
Assuming an X--ray to optical spectral index appropriate for these
luminosities ($\alpha_{\rm ox}=-1.65$, e.g.Vignali et al. 2003, Steffen et
al. 2006), the absolute magnitude SDSS limit corresponds to an X--ray
luminosity  log$L_X \sim 45$. 
Therefore, the observed decline in the space density for the
XMM--COSMOS sample strongly suggests that the evolution of mildly obscured
(Compton thin) AGN is very similar to that of unobscured, optically luminous
QSOs, provided that the shape of the declining function holds also for luminosities 
which are about an order of magnitude lower than those probed by SDSS.  

Silverman et al. (2008) computed the XLF for the optically bright
($i<24$) X-ray population. 
Even though there is a good agreement at $z <$ 3.5 between their 
LDDE XLF and the observed XMM-COSMOS space densities 
(for the same limit in the optical magnitude), 
the extrapolation of their XLF at higher redshift is larger, by a factor of
$\sim2$,   
with respect to our data. A better fit would be obtained by 
tuning the Silverman et al. (2008) LF parameters responsible for the high 
redshift ($z > $ 3.0--3.5) behavior.


The observed decline in the comoving space density of X--ray selected QSOs, 
at least for luminosities larger than $\sim10^{44}$ erg s$^{-1}$,
has a significant impact on the predictions of the QSOs number counts 
expected from future large area X--ray  surveys, and, more in general, on
predictions 
at all the wavelengths based on the current available XLF as representative of
the high-redshift, radio quiet population (e.g. Wilman et al. 2008).
The present results may also provide a benchmark for theoretical models 
of SMBH growth. For example, the $z >$ 4 QSOs number counts predicted by 
the Rhook \& Haehnelt (2008) models (see central and right panels in their
Fig.~6) are about a factor 7--8 higher than what is actually observed,
suggesting that some of the model parameters should be revised. 

The predicted number counts for a model with and without an exponential
decline, at different limiting fluxes and redshift ranges are listed in table
4. The Chandra-COSMOS survey provides a homogeneous coverage with
high-resolution (HPD $<2''$) over the central $\sim0.5$ deg$^2$ in COSMOS
(Elvis et al. 2008) down to a limiting flux of $\sim4\times10^{-16}$ 
\cgs\ in the 0.5-2 keV band (see vertical line in Fig.~8). It will therefore
roughly double the number of $z>3$ quasars in that area, exploring the flux
regime where the contribution from most obscured sources is higher. 

As a practical example we also report the expectations for the eROSITA 
(extended ROentgen Survey with an Imaging Telescope Array) survey 
with the SRG (Spectrum R\"ontgen Gamma) mission , which will be
launched in the next years. eROSITA will survey the entire extragalactic sky
down to a limiting flux of $\sim 10^{-14}$ \cgs in the 0.5-2 keV band. The
number of expected $z>3$ ($z>4$) QSOs in the all sky survey, when the decline
described in $\S$5.2 is incorporated in our current knowledge of the XLF, is
2.5$\times10^4$ (2100).  eROSITA will also perform a deeper survey over an area
of $\sim 400$ deg$^2$ down to a depth of $4\times10^{-15}$ \cgs.  We expect to
reveal $\sim 2500$ QSOs at $z>3$, about 200 of them at $z>4$ and better
constrain their redshift distribution.

The study of QSOs space density and evolution at lower X--ray luminosities 
and higher redshifts requires much deeper observations. 
While a few moderately luminous (log$L_X \sim$ 43--44), high redshift 
$z >$ 4 quasars are detected in deep {\it Chandra} fields (i.e. Vignali et 
al. 2002) the present sample size is not such to constrain their space density.
Additional high--z objects are expected to be revealed by 
ongoing  ultra-deep {\it Chandra} (Luo et al. 2008) and
XMM--{\it Newton} surveys.

\section{Summary}

Taking advantage of the large area of the XMM--COSMOS survey and the associated
deep multiwavelength follow-up, we have studied the physical and cosmological 
properties of $z >$ 3, X--ray selected QSOs. This sample of 40 objects 
constitutes the largest and most complete (in term of spectroscopic
confirmation, 55\%) sample of high redshift quasars published so far and
extracted from a single survey, at the depth of $10^{-15}$ \cgs\ in the 0.5-2 keV band. 

The most important results of our analysis can be summarized as follows: 
\begin{itemize} 
\item  X--ray selected QSOs have optical properties which are  
       not significantly different from those of optically selected
       (i.e. SDSS) objects. 
\item   From the analysis of X--ray colors and stacked spectra 
        there is evidence of substantial X--ray absorption in about 20\% of the sources.
\item  There is no clear correlation between X--ray obscured sources and 
       outliers in the adopted optical color--color diagram; however, 
       the outliers may contribute to the flatter X--ray spectrum observed
       for the spectroscopically unidentified objects sample for which the
       observed HR does not allow to constrain the column density.      
\item   
   A steep decline in the space density of luminous X--ray selected QSOs
  ($L_X >$ 10$^{44.2}$ erg s$^{-1}$) at $z>$ 3, similar to that observed for
  luminous optically selected quasars, is needed to fit the observed data. 
  This suggests that the evolution of obscured AGN is similar to that of
  unobscured ones.  
\end{itemize}
We emphasize that, while the currently available XLF (e.g. Ueda et al. 2003, La 
Franca et al. 2005, Hasinger et al. 2005, Silverman et al. 2008) provide an
excellent fit to the $z<3$ redshift population, their extrapolations
to high redshift would overpredict the present observational constraints by a
factor $\sim 2$ at $z \sim3$ and $\gsimeq5$ at $z\sim4$. As far as the
luminous quasars are concerned (logL$_X>44.2$), a decay in their space density
at $z\gsimeq3$ should be included in the predictions of the high--z 
quasar population. 

Future medium deep all sky X-ray surveys (e.g. eROSITA) with the associated 
multiwavelength follow--up will provide sizable samples ($>10^{4}$) of
$z>3$ QSOs to further constrain the bright end of the QSOs luminosity
function. The evolution of the faint end of
X--ray selected QSOs luminosity function up to very high redshift,
$z \sim$ 6 and perhaps beyond, will be investigated by future missions 
such as IXO and Generation--X.

\acknowledgments
This work is based on observations obtained with XMM-{\it Newton}, 
an ESA Science Mission with instruments
and contributions directly funded by ESA Member States and the
USA (NASA). In Germany, the XMM-{\it Newton} project is supported by the
Bundesministerium f\"ur Wirtschaft und Technologie/Deutsches Zentrum
f\"ur Luft- und Raumfahrt (BMWI/DLR, FKZ 50 OX 0001), and the Max-Planck
Society. MB acknowledge support from the XMM--{\it Newton} DLP grant
50-)G-0502, GH ackwnoledge contribution from the Leibniz Prize of the
Deutsche Forschungsgemeinschaft under the grant HA 1850/28-1. 
In Italy, the XMM-COSMOS project is supported by ASI-INAF and PRIN/MIUR under
grants I/023/05/00 and 2006--02--5203. 
The zCOSMOS ESO Large Program Number 175.A-0839 is acknowledged.
These results are also based in part on observations obtained with the
Walter Baade telescope of the Magellan Consortium at Las Campanas Observatory, 
and with the MMT which is operated by the MMT Observatory (MMTO), a joint
venture of the Smithsonian Institution and the University of Arizona.
We gratefully acknowledge the contributions of the entire COSMOS
collaboration; more information on the COSMOS survey is available at 
\verb+http://www.astro.caltech.edu/~cosmos+. This research has made  
use of the NASA/IPAC Extragalactic Database (NED) and the SDSS spectral 
archive. 
 We thank the anonymous referee for his/her useful comments and suggestions.

%

\begin{deluxetable}{rrrcc}
\tablecaption{List and coordinates of the $z>3$ QSOs from the XMM-COSMOS sample}
\tablehead{
\colhead{IAU name$^a$} & \colhead{XID$^b$} & \colhead{ID$^c$} & \colhead{RA$^d$} &
\colhead{DEC$^d$} \\
    &       &       &  hh:mm:ss & dd:mm:ss }
\startdata
XMMC\_J100101.6+023846            &         326 & 2668917 &    10:01:01.51  & +02:38:48.68 \\
XMMC\_J100157.5+014446            &        2518 &  448852 &    10:01:57.72  & +01:44:47.05 \\
XMMC\_J100119.9+023444            &         418 & 2289051 &    10:01:20.05  & +02:34:43.82 \\
XMMC\_J100226.0+024611            &        5331 & 2558690 &    10:02:26.11  & +02:46:10.78 \\
XMMC\_J100223.2+022557            &       10690 & 1849901 &    10:02:23.31  & +02:25:58.17 \\
XMMC\_J100050.6+022328            &         349 & 1977373 &    10:00:50.58  & +02:23:29.30 \\
XMMC\_J095755.4+022400            &        2407 & 2132709 &    09:57:55.47  & +02:24:01.18 \\
XMMC\_J095931.8+023018            &         262 & 2421306 &    09:59:31.80  & +02:30:18.51 \\
XMMC\_J095806.9+022248            &        2421 & 2137194 &    09:58:06.98  & +02:22:48.59 \\
XMMC\_J095840.6+021003            &        5347 & 1749560 &    09:58:40.71  & +02:10:03.72 \\
XMMC\_J095859.5+024356            &        5161 & 2768843 &    09:58:59.70  & +02:43:55.25 \\
XMMC\_J100228.8+024017            &        5175 & 2583306 &    10:02:28.82  & +02:40:17.08 \\
XMMC\_J100220.3+020452            &        5345 & 1131048 &    10:02:20.37  & +02:04:52.91 \\
XMMC\_J100023.4+020115            &         469 & 1259457 &    10:00:23.24  & +02:01:17.40 \\
XMMC\_J100256.8+024320            &        5219 & 2534376 &    10:02:56.92  & +02:43:21.27 \\
XMMC\_J095740.6+025259            &       60311 & 3190185 &    09:57:40.70  & +02:52:58.77 \\
XMMC\_J100113.3+014542            &       53733 &  485583 &    10:01:13.45  & +01:45:41.89 \\
XMMC\_J100111.4+020853            &       60131 & 1593500 &    10:01:11.34  & +02:08:55.84 \\
XMMC\_J100050.1+022855            &         180 & 2350265 &    10:00:50.12  & +02:28:54.97 \\
XMMC\_J100127.5+020837            &        2394 & 1594618 &    10:01:27.53  & +02:08:37.79 \\
XMMC\_J095928.7+021738            &        1151 & 2039436 &    09:59:28.72  & +02:17:38.57 \\
XMMC\_J100057.8+023931            &         187 & 2665989 &    10:00:57.79  & +02:39:32.61 \\
XMMC\_J095923.0+022853            &        5482 & 2426654 &    09:59:22.98  & +02:28:54.03 \\
XMMC\_J100255.6+013057            &       60186 &   46132 &    10:02:55.81  & +01:30:58.05 \\
XMMC\_J100145.6+024212            &        5382 & 2615665 &    10:01:45.58  & +02:42:12.59 \\
XMMC\_J100256.6+021159            &        5116 & 1462117 &    10:02:56.53  & +02:11:58.48 \\
XMMC\_J100000.9+020220            &        5583 & 1294973 &    10:00:01.05  & +02:02:20.03 \\
XMMC\_J095901.2+024419            &        5162 & 2767217 &    09:59:01.29  & +02:44:18.81 \\
XMMC\_J100312.0+024916            &       53351 & 2910750 &    10:03:12.06  & +02:49:15.75 \\
XMMC\_J095753.2+024737            &        5199 & 3176366 &    09:57:53.49  & +02:47:36.25 \\
XMMC\_J095854.2+023753            &        5525 & 2832144 &    09:58:54.36  & +02:37:53.38 \\
XMMC\_J100232.9+022331            &       60007 & 1859446 &    10:02:33.23  & +02:23:28.86 \\
XMMC\_J095931.0+021332            &         504 & 1694357 &    09:59:31.01  & +02:13:33.01 \\
XMMC\_J100104.2+014202            &        2602 &  500928 &    10:01:04.17  & +01:42:03.13 \\
XMMC\_J100050.2+022618            &         300 & 1965822 &    10:00:50.16  & +02:26:18.48 \\
XMMC\_J100248.9+022210            &        5592 & 1864254 &    10:02:48.90  & +02:22:12.07 \\
XMMC\_J095906.4+022638            &        5606 & 2042408 &    09:59:06.46  & +02:26:39.50 \\
XMMC\_J095752.1+015118            &        5594 & 1061300 &    09:57:52.16  & +01:51:20.15 \\
XMMC\_J095856.7+021047            &       54439 & 1705273 &    09:58:56.69  & +02:10:47.75 \\
XMMC\_J100152.0+023152            &        5259 & 2260872 &    10:01:52.10  & +02:31:55.19 \\

\hline
\enddata\\
\tablenotetext{a}{Official IAU designation for the XMM-COSMOS sources}
\tablenotetext{b}{X--ray identifier number from the XMM-COSMOS catalog
(Cappelluti et al. 2008)}
\tablenotetext{c}{Optical identifier number from the photometric COSMOS
  catalog published by Capak et al. (2007). Photometry in the optical bands 
can be retrieved from IRSA at the link: 
http://irsa.ipac.caltech.edu/data/COSMOS/tables/cosmos\_phot\_20060103.tbl.gz}
\tablenotetext{d}{Coordinates of the optical counterpart of the X--ray source}
\end{deluxetable}

\begin{deluxetable}{rcccccr}
\tablecaption{Properties of $z>3$ QSOs from the XMM-COSMOS sample}
\tablehead{
\colhead{XID} & \colhead{zspec} & \colhead{zphot} & \colhead{i (AB)} & 
  \colhead{F$_{0.5-2~\rm keV}$}  & \colhead{logL$_{2-10~\rm keV}$$^h$} 
  & \colhead{HR} \\  
    &       &       &   & 10$^{-15}$ \cgs\ &  erg s$^{-1}$ & }
\startdata
     326 &    3.003$^c$ &    3.06 (3.04-3.08) &  23.23 & 1.21  & 44.05 & $<-0.53$ \\ 
    2518 &      ... &    3.01 (2.96-3.06)$^a$  &  23.96     & 1.31  & 44.26 &  0.17 \\ 
     418 &      ... &    3.03 (1.58-3.84)$^{b,e}$  &  27.02 & 1.73  & 44.25 & -0.28 \\ 
    5331 &    3.038 &    3.04 (3.02-3.06) &  20.28  & 4.58  & 44.62 & -0.59 \\ 
   10690 &      ... &    3.09 (3.06-3.12) &  23.24 & 1.29  & 44.10 & $<-0.40$ \\ 
     349 &    3.092 &  3.09 (3.06-3.10) &  22.58 & 1.76  & 44.29 & -0.23 \\ 
    2407 &    3.093 &      0.99 (0.96-1.02) &  21.26 & 11.4 & 45.06 & -0.46 \\
     262 &      ... &    3.10 (2.46-3.26)$^{a,e}$ &  25.82 & 5.69  & 44.75 & -0.50 \\ 
    2421 &    3.104 &     3.11 (3.08-3.12) &  20.89  & 5.22  & 44.73 & -0.41 \\ 
    5347 &      ... &    3.13 (2.60-3.24)$^b$  &  24.71 & 2.05  & 44.29 & $<-0.57$ \\ 
    5161 &      ... &    3.14 (3.10-3.18)  &  24.00 & 1.64  & 44.29 & -0.14 \\ 
    5175 &    3.143 &    3.13 (3.10-3.16) &  22.07  & 6.63  & 44.84 & -0.44 \\ 
\hline
    5345 &      ... &    3.29 (3.26-3.32)  &  23.15 & 1.86  & 44.31 & $<-0.50$ \\ 
     469 &      ... &    3.30 (3.14-3.40)  &  24.42 & 1.08  & 44.10 & $<-0.40$ \\ 
    5219 &    3.304 &     3.25 (3.22-3.28) &  22.25 & 2.21  & 44.39 & $<-0.42$ \\ 
   60311 &      ... &    3.31 (3.28-3.34)  &  21.92 & 2.99  & 44.51 & $<-0.11$ \\ 
   53733 &      ... &    3.32 (3.26-3.40)  &  25.02 & 1.00  & 44.07 & $<-0.44$ \\ 
   60131 &    3.328$^{d}$ & 3.47 (3.44-3.50) &  20.53 & 1.09  &  44.21 & 0.10 \\ 
     180 &    3.333 &     3.34 (3.32-3.36) &  20.17 & 2.96  & 44.51 & $<-0.73$ \\ 
    2394 &    3.333$^c$ &    3.31 (3.28-3.34) &  23.16 & 3.37  & 44.61 & -0.36 \\
    1151 &    3.345$^{d}$ &  3.38 (3.36-3.40) & 21.77 & 3.41  & 44.67 & -0.08 \\ 
     187 &    3.356$^{c}$ & 3.33 (3.30-3.36) & 22.36  & 3.78  & 44.67 & -0.38 \\ 
    5482 &      ... &    3.36 (3.34-3.38)  &  23.22  & 1.70  & 44.34 & -0.26 \\ 
   60186 &      ... &    3.40 (0.22-3.42) $^b$  &  22.11 & 2.98  &  44.54 & $<-0.08$ \\ 
\hline
    5382 &    3.465 &     3.44 (3.42-3.46)  &  23.08  & 1.14  & 44.17 & $<-0.31$ \\ 
    5116 &      ... &    3.47  (3.46-3.50) &  20.29   & 4.15  & 44.73 & -0.43 \\ 
    5583 &    3.499 &     3.49  (3.48-3.52) &  21.88  & 1.48  & 44.28 & $<-0.57$ \\ 
    5162 &    3.524$^{e,g}$ & 3.52 (3.50-3.56) & 22.93 & 1.92  & 44.45 & -0.11 \\ 
   53351 &      ... &    3.53 (2.22-3.70) $^{b,e}$  &  24.86 & 2.84  & 44.63 & -0.04 \\ 
    5199 &    3.609 &      3.61 (3.58-3.62) &  21.96 & 2.68  & 44.60 & -0.21 \\ 
    5525 &      ... &    3.65 (3.62-3.66) &  22.28  & 2.82  & 44.64 & -0.13 \\ 
   60007 &     ...  &    3.65 (3.62-3.68) &  22.17  & 0.96$^{f}$ &  44.21 & 0.14  \\ 
     504 &    3.651 &      3.65 (3.62-3.66) &  22.72 & 1.48  & 44.32 & $<-0.53$ \\ 
    2602 &      ... &    3.70 (1.42-4.04)$^b$  &  24.79 & 1.87  & 44.44 & $<-0.62$ \\ 
     300 &    3.715 &    3.54 (3.52-3.56) &  21.24  & 1.62  & 44.38 & $<-0.57$ \\ 
    5592 &    3.745$^c$ &    3.76 (3.74-3.80)  &  22.47 & 2.21  & 44.52 & $<-0.41$ \\ 
    5606 &    4.166$^{e,g}$  &    4.01 (3.96-4.06) &  22.79  & 1.33  & 44.42 & $<-0.51$ \\ 
    5594 &    4.174 &    3.73 (3.70-3.74) &  21.08  & 1.27  & 44.40 & $<-0.28$ \\ 
   54439 &    4.241$^{e}$ &  4.24 (4.22-4.26) &  23.53 & 1.36  & 44.45 & $<-0.37$ \\ 
    5259 &      ... &    4.45 (4.42-4.50)$^{e}$  &  26.36 & 1.08  & 44.39 & $<-0.49$ \\ 
\hline
\enddata\\
\tablenotetext{h}{Rest-frame hard X--ray luminosity}
\tablenotetext{a}{Objects with large $\chi^2$ minima (extending at
  z$<3$)}
\tablenotetext{b}{Objects which have a comparable photoz solution at z$<3$}
\tablenotetext{c}{Single line spectrum}
\tablenotetext{d}{BAL QSO}
\tablenotetext{e}{Undetected in the U-band}
\tablenotetext{f}{Source detected in the hard band, but below the threshold
  (10$^{-15}$ \cgs\ ) in the soft band}
\tablenotetext{g}{Candidate NL QSO (FWHM $<1500$ km s$^{-1}$)}

\end{deluxetable}

\begin{center}
\begin{deluxetable}{lrrc}
\tablecaption{Summary of stacking results for the $z>3$ QSOs samples}
\tablehead{
\colhead{sample} & \colhead{\# of sources} & \colhead{Total net counts} & 
\colhead{$\Gamma$} \\ 
&  &  &  simple PL } 
\startdata
log $N_H <$ 23 & 26 & 1200.9 & 1.8$^{+0.2}_{-0.3}$ \\ 
\hline
log $N_H >$ 23 & 9 & 342.4 & 0.9$^{+0.7}_{-0.8}$ \\
\hline
log $N_H <$ 23, BL AGN & 16 & 782.1   & 1.7$^{+0.3}_{-0.3}$ \\
\hline
log $N_H <$ 23, phot-z & 10 & 418.8 & 1.5$^{+0.4}_{-0.5}$ \\
\enddata
\end{deluxetable}
\end{center}

\begin{center}
\begin{deluxetable}{rlcc}
\tablecaption{Expected numbers of z$>3$ QSOs in 0.5-2 keV surveys}
\tablehead{
\colhead{z range} & \colhead{limiting flux} & \colhead{constant$^a$} &
\colhead{decline$^b$} \\ 
                  & \cgs\ &  deg$^{-2}$ & deg$^{-2}$}
\startdata
$z>3$ & $>4\times10^{-16}$ & 230 &  75 \\ 
      & $>10^{-15}$ & 80 &  30 \\ 
      & $>4\times10^{-15}$ &  14 &  6.2 \\ 
      & $>10^{-14}$ &  1.8  &  0.75 \\ 
\hline
$z>4$ & $>4\times10^{-16}$ & 80 &  12 \\ 
      & $>10^{-15}$ & 30 &  7 \\ 
      & $>4\times10^{-15}$ & 3 &  0.5 \\ 
      & $>10^{-14}$ & 0.6  &  0.07 \\ 
\enddata
\tablenotetext{a}{Predicted number counts assuming extrapolating the Hasinger
  et al. (2005) XLF at high redshifts} 
\tablenotetext{b}{Predicted number counts assuming the Hasinger et al. (2005)
  XLF modified by an exponential decline at $z>2.7$} 
\end{deluxetable}
\end{center}

\end{document}